%% file: book.tex
\documentclass[fleqn]{book}

\def\gapprox{\lower.4ex\hbox{$\;\buildrel >\over{\scriptstyle\sim}\;$}}
\def\lapprox{\lower.4ex\hbox{$\;\buildrel <\over{\scriptstyle\sim}\;$}}

\def\ref#1{\par\noindent\hangindent1cm {#1}}

\def\captio#1{\caption{\small {#1} \normalsize}}

\usepackage{graphics}
\usepackage{graphicx}
\usepackage{epsfig}
\usepackage{times}
\usepackage{makeidx}

\begin{document}
\renewcommand{\topfraction}{0.95}
\renewcommand{\bottomfraction}{0.95}
\renewcommand{\textfraction}{0.05}
\renewcommand{\floatpagefraction}{0.95}
\renewcommand{\dbltopfraction}{0.95}
\renewcommand{\dblfloatpagefraction}{0.95}
\newcommand{\AaA}{Astron.~Astrophys.\ } 
\newcommand{\ApJ}{Astrophys.~J.\ }      
\newcommand{\GRL}{Geophys.~Res.~Lett.\ }
\newcommand{\JGR}{J.~Geophys.~Res.\ }   
\newcommand{\SP}{Solar~Phys.\ }         

\setcounter{tocdepth}{3}

 \frontmatter
 \tableofcontents\cleardoublepage
 \mainmatter
 \include{c02}



\end{document}

%% file: c02.tex
\setcounter{chapter}{1} 
\setcounter{page}{1}

\chapter{Theoretical Models of SOC Systems}

\subsection*{by Markus J. Aschwanden}

\bigskip
{\sl How can the universe start with a few types of elementary particles
at the big bang, and end up with life, history, economics, and
literature? The question is screaming out to be answered but it is seldom
even asked. Why did the big bang not form a simple gas of particles, or
condense into one big crystal?} (Bak 1996). The answer to this fundamental
question lies in the tendency of the universal evolution towards complexity,
which is a property of many nonlinear energy dissipation processes.
Dissipative nonlinear systems generally have a source of free energy,
which can be partially dissipated whenever an instability occurs. This
triggers an avalanche-like energy dissipation event above some threshold
level. Such nonlinear processes are observed in astrophysics, magnetospheric
physics, geophysics, material sciences, physical laboratories,
human activities (stock market, city sizes, internet, brain activity),
and in natural hazards and catastrophes (earthquakes,
snow avalanches, forest fires). A tentative list of SOC phenomena with 
the relevant sources of free energy, the physical driver mechanisms, and 
instabilities that trigger a SOC event are listed in Table 2.1.

A prominent theory that explains such nonlinear energy dissipation events
is the so-called {\sl Self-organized criticality (SOC)} concept, first
pioneered by Bak et al. (1987, 1988) and simulated with cellular
automaton models, which mimic next-neighbour interactions leading
to complex patterns. The topic of SOC is reviewed in recent reviews,
textbooks, and monographs (e.g., Bak 1996; Jensen 1998; Turcotte 1999;
Charbonneau et al.~2001; Hergarten 2002; Sornette 2004; Aschwanden 2011a;
Crosby 2011; Pruessner 2012).
SOC can be considered as a basic physics phenomenon - universally occurring
in systems with many coupled degrees of freedom in the limit of infinitesimal
external forcing.
This theory assumes a critical state that is
robust in the sense that it is self-organizing, like a critical slope of a
sandpile is maintained under the steady (but random) dropping of new
sand grains on top of the pile. Individual avalanches occur with
unpredictable sizes, uncorrelated to the disturbances produced by the
input. Sandpile avalanches are a paradigm of the SOC theory, which
has the following characteristics: (1) Individual events are statistically
independent, spatially and temporally (leading to random
waiting time distributions); (2) The size
or occurrence frequency distribution is scale-free and can be
characterized by a powerlaw function over some
size range; (3) The detailed spatial and temporal evolution is complex 
and involves a fractal geometry and stochastically fluctuating (intermittent) 
time characteristics (sometimes modeled
with 1/f-noise, white, pink, red, or black noise). 

\begin{table}
\begin{center}
\normalsize
\caption{Examples of physical processes with SOC behavior.}
\medskip
\begin{tabular}{|l|l|l|}
\hline
SOC Phenomenon    &Source of free energy        &Instability or         \\
                  &or physical mechanism        &trigger of SOC event   \\
\hline
\hline
Galaxy formation  &gravity, rotation            &density fluctuations   \\
Star formation    &gravity, rotation            &gravitational collapse \\
Blazars           &gravity, magnetic field      &relativistic jets      \\
Soft gamma ray repeaters &magnetic field        &star crust fractures   \\
Pulsar glitches   &rotation                     &Magnus force           \\
Blackhole objects &gravity, rotation            &accretion disk instability\\
Cosmic rays       &magnetic field, shocks       &particle acceleration  \\
Solar/stellar dynamo&magnetofriction in tachocline&magnetic buoyancy    \\
Solar/stellar flares&magnetic stressing         &magnetic reconnection  \\
Nuclear burning   &atomic energy                &chain reaction         \\
Saturn rings      &kinetic energy               &collisions             \\
Asteroid belt     &kinetic energy               &collisions             \\
Lunar craters     &lunar gravity                &meteroid impact        \\
Magnetospheric substorms &electric currents, solar wind &magnetic reconnection\\
Earthquakes       &continental drift            &tectonic slipping      \\
Snow avalanches &gravity                        &temperature increase   \\
Sandpile avalanches&gravity                     &super-critical slope   \\
Forest fire       &heat capacity of wood        &lightening, campfire   \\
Lightening        &electrostatic potential      &discharge              \\
Traffic collisions&kinetic energy of cars       &driver distraction, ice\\
Stockmarket crash &economic capital, profit     &political event, speculation\\
Lottery win       &optimistic buyers            &random drawing system  \\
\hline
\end{tabular}
\end{center}
\end{table}

There are some
related physical processes that share some of these characteristics,
and thus are difficult to discriminate from a SOC process, such as
turbulence, Brownian motion, percolation, or chaotic systems (Fig.~2.1,
right column).
A universal SOC theory that makes quantitative predictions of the
powerlaw-like occurrence frequency and waiting time distributions
is still lacking. We expect that the analysis of large new
databases of space observations and geophysics records (available
over at least a half century now) will provide unprecedented statistics 
of SOC observables (Fig.~2.1, left column), which will constrain the observed
and theoretically predicted statistical distribution functions
(Fig.~2.2, middle column), and this way confirm or disprove the
theoretical predictions of existing SOC theories and models (Fig.~2.2, right
column). Ultimately we expect to find a set of observables that allows us 
to discriminate among different SOC models as well as against 
other nonlinear dissipative processes (e.g., MHD turbulence).
In the following we will describe existing SOC models and
SOC-related processes from a theoretical point of view.

\begin{figure}
\centerline{\includegraphics[width=1.0\textwidth]{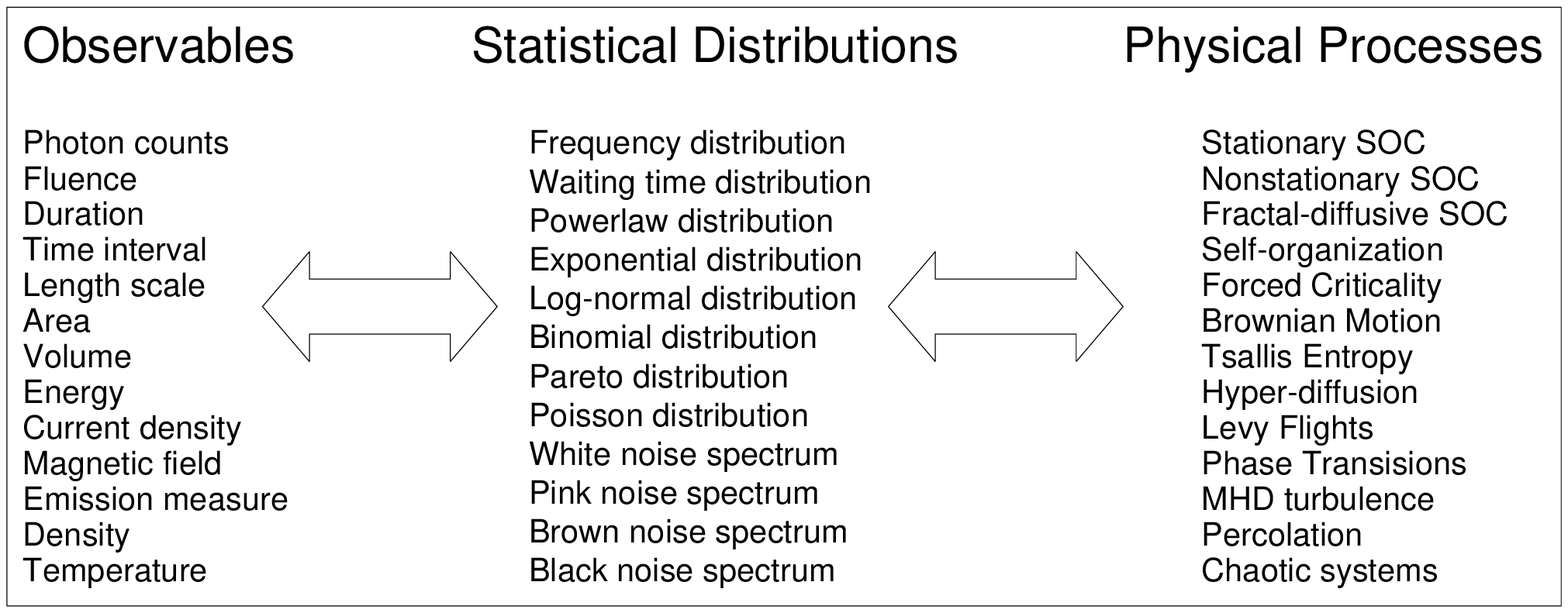}}
\caption{Metrics of observables, statistical distributions, and physical
models that need to be defined in order to discriminate SOC from non-SOC
processes.}
\end{figure}

\section{Cellular Automaton Models (CA-SOC)} 

The theoretical models can be subdivided into: (i) a mathematical part
that deals with the statistical aspects of complexity, which is universal
to all SOC phenomena and essentially ``physics-free'' (such as the
powerlaw-like distribution functions), and (ii) a physical part that links 
the observable parameters to a particular physical mechanism (for instance
in form of scaling law relationships between physical parameters).

\subsection{Statistical Aspects} 

The original concept of self-organized criticality was pioneered by
Bak et al.~(1987, 1988), who used the paradigm of a sandpile with a
critical slope to qualitatively illustrate the principle of self-organized
criticality. The first theoretical description of the SOC behavior of
a nonlinear system was then studied with a large number of coupled
pendulums and with numerical experiments that have been dubbed
{\sl cellular automaton} models (CA-SOC). In essence, a SOC system
has a large number of elements with coupled degrees of freedom, where
a random disturbance causes a complex dynamical spatio-temporal pattern.
Although such a nonlinear system obeys classical or quantum-mechanical
physics, it cannot be described analytically because of the large number
of degrees of freedom. The situation is similar to the N-body problem,
which cannot be solved analytically for more complex configurations than
a two-body system. However, such SOC systems can easily be simulated
with a numerical computer program, which starts with an initial condition
defined in a lattice grid, and then iteratively updates the dynamical 
state of each system node, and this way mimics the dynamical
evolution and statistical distribution functions of SOC parameters.
Hence, such numerical lattice simulations of SOC behavior (i.e., 
cellular automatons) were the first viable method to study the SOC 
phenomenon on a theoretical basis. 

What are the ingredients and physical parameters of a cellular automaton 
model? What is common to all cellular automaton models of the type of
Bak, Tang, and Wiesenfeld (1987, 1988), briefly called BTW models, is:
(1) A S-dimensional rectangular lattice grid, say a coordinate system
$x_{i,j,k}, i=1,...,n; j=1,...,n, k=1,...,n$ for a 3-dimensional case
$S=3$ with grid size $n$; (2) a place-holder for 
a physical quantity $z_{i,j,k}$ associated with each cellular node
$x_{i,j,k}$, (3) a definition of a critical threshold $z_{crit}$,
(4) a random input $\Delta z_{i,j,k}$ in space and time; (5) a
mathematical re-distribution rule that is applied when a local physical
quantity exceeds the critical threshold value, for instance a critical slope
of a sandpile, which adjusts the state of the next-neighbor cells, i.e.,
$$
        \begin{array}{ll}
        z_{i,j,k}=z_{i,j,k}+1 & {\rm initial\ input} \\
        z_{i,j,k}=z_{i,j,k}-8 & {\rm if}\ z_{i,j,k} \ge 8 , \\
        z_{i\pm 1,j\pm 1,k\pm 1}=z_{i\pm 1,j\pm 1,k\pm 1}+1 &
        \end{array}
        \eqno(2.1)
$$
for the 8 next neighbors in a 3D-lattice grid, and (6) iterative steps
in time to update the system state $z_{i,j,k}(t)$ as a function of time $t$.
Such cellular automaton simulations usually start with a stable initial
condition (e.g., an empty system with $z_{i,j,k}(t=0)=0$), and have then
to run several million time steps before they reach a critical state.
Once they reach the critical state, avalanches of arbitrary sizes can
be triggered by the random input of an infinitesimal disturbance
$\Delta z_{i,j,k}(t)$, separated by quiescent time intervals in between,
in the case of slow driving. Statistics of the occurrence frequency
distributions of avalanche sizes or durations yields then the ubiquitous 
powerlaw-like distribution functions. 
Fig.~2.2 shows an example of a 2D-lattice simulation, displaying a
snapshot of a large avalanche (top left panel) as well as the statistical
distribution function of avalanche sizes $L$ (left) and durations $T$
(right panels). 

\begin{figure}
\centerline{\includegraphics[width=1.0\textwidth]{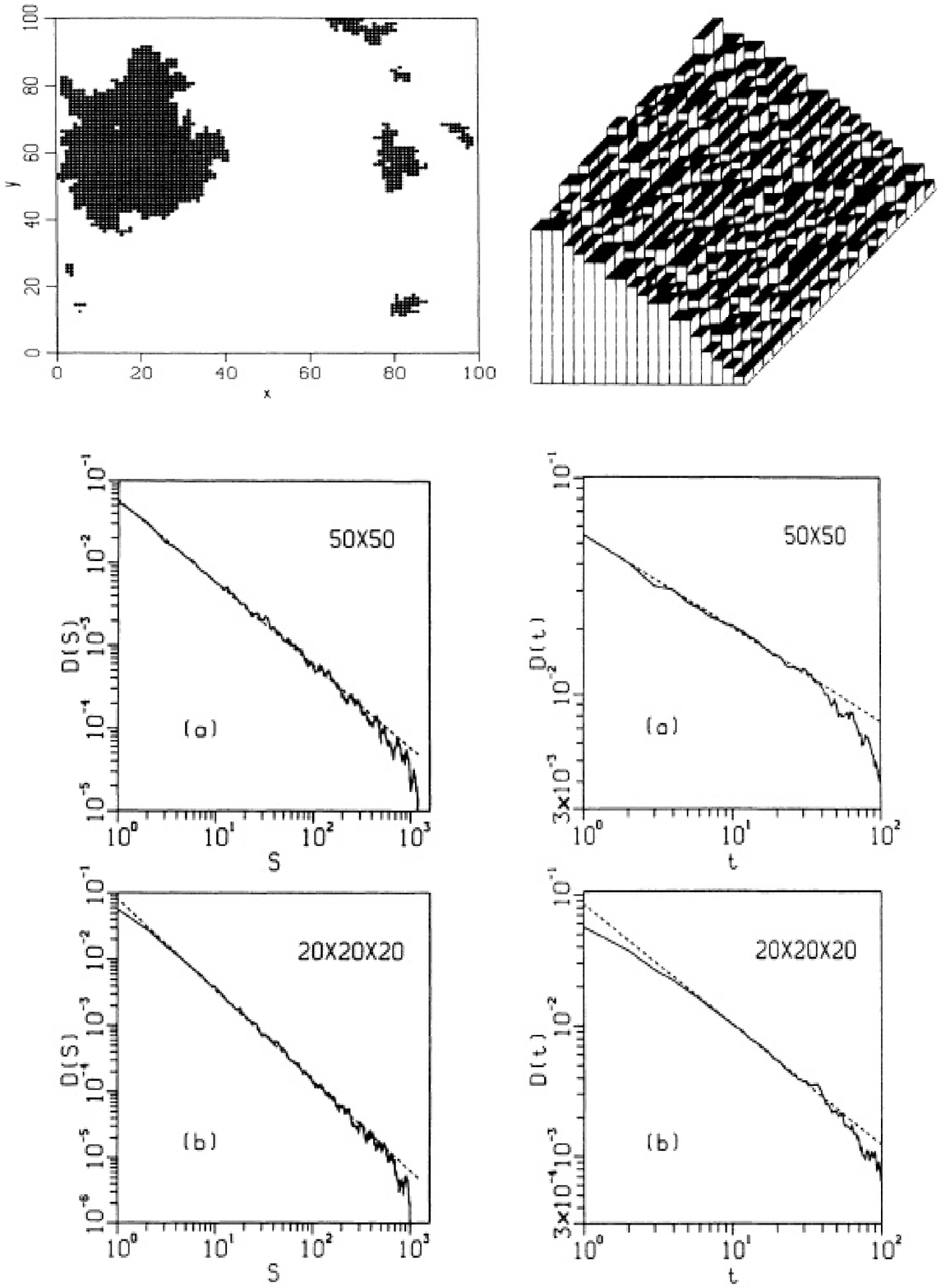}}
\captio{Examples of a fragmented avalanche (top left) occurring in a
2-D (computer) sandpile (top right) and occurrence
frequency distribution of avalanche cluster sizes (left panels) and
avalanche durations (right panels) of the original BTW sandpile
cellular automaton simulation. The simulations have been performed
for a 50$\times$50 2-D lattice (middle panels) and for
a 20$\times$20$\times$20 3-D lattice grid (bottom panels).
The powerlaw slopes are $\alpha_S=1.0$ and $\alpha_T=0.42$ for the 2-D grid
(middle panels) and $\alpha_S=1.37$ and $\alpha_T=0.92$ for the 3-D grid.
Reprinted from Bak, Tang, and Wiesenfeld (1987, 1988)   
with permission; Copyright by American Physical Society.}
\end{figure}

Thus, such SOC cellular automaton models primarily quantify the 
spatio-temporal dynamics of complex systems and the related statistics 
of emerging spatio-temporal patterns. They can predict the distribution 
function $N(L) dL$ of the size or length scale $L$ of dynamical events 
(also called ``avalanches''), or the time scale $T$ of an avalanche event. 
Secondary dynamical parameters can be derived, such as the instantaneous 
avalanche volume $V(t)$ or its change $dV(t)/dt$ during an avalanche event, 
or the total time-integrated volume of the avalanche $\int_0^T V(t) dt$. 
However, such SOC cellular automaton models are ``physics-free'', because
the dynamics of an event is described by a mathematical rule that substitutes
for the unknown physical process, but is thought to be an universal property
of SOC systems in a critical state. Applying cellular automaton simulations
to observed phenomena requires than to substitute a physical quantity into
the ``place-holder'' variable $z_{i,j,k}$, which is otherwise undefined
in cellular automata, except for defining the instability criterion
$z_{i,j,k} > z_{crit}$ in terms of a critical threshold $z_{crit}$.  

There is a large industry of cellular automaton models, mostly 
modified versions of the original BTW-model, such as the lattice-gas
model (Jensen 1998), Conway's game of life model (Bak, Chen and Creutz 1989),
traffic jam simulations (Nagel and Paczuski 1995), multiple-strategy
agent-based models applied to the financial market (Feigenbaum 2003),
punctuated equilibrium models applied to biophysics (Bak and Sneppen 1993), 
slider-block spring models applied in geophysics (Burridge and Knopoff 1967),
forest-fire models (Malamud et al.~1998; reviewed in Turcotte 1999),
applications to magnetospheric substorms (Takalo et al.~1993; 1999a,b),
to solar flares (Lu and Hamilton 1991; Charbonneau et al.~2001),
and to stellar accretion disks (Mineshige et al.~1994a,b; 
Pavlidou et al.~2001). 
More complete reviews of such cellular automaton models are given in 
Turcotte (1999) and Aschwanden (2011a, Section 2).

\subsection{Physical Aspects}

Every theoretical model needs verification by experiments and observations.
Since a cellular automaton model is a purely mathematical model of
complexity, similar to the mathematical definition of a fractal dimension, 
its predictions of a particular mathematical distribution function 
(such as a powerlaw) may represents a universal property of SOC
processes, but does not represent a complete physical model {\sl per se} 
that can be used for applications to real observations. The physical
aspect of a SOC model is hidden in the ``place-holder'' variable 
$z_{i,j,k}$ and its scaling laws with observables, which require a 
specific physical mechanism for each observed phenomenon (for examples 
see second column in Table 2.1). 

If we go back to the original SOC concept of avalanches in a sandpile,
say in the 1D-version, the place-holder variable $z_i$ has been interpreted
as a vertical altitude or height difference $z_i=h_{i+1}-h_i$ between
next-neighbor nodes, so that the instability threshold corresponds to
a critical slope $z_i > z_{crit} = (dh/dx)_{crit}$. For the 3D version,
the re-distribution rule is given in Eq.~(2.1). The instability 
threshold essentially corresponds to a critical point where the
gravitational force exceeds the frictional force of a sand grain. 

In applications to solar flares or magnetospheric substorms, 
the ``place-holder'' variable $z_{i,j,k}$
has been related to the magnetic field $B_{i,j,k}=B(x_{i,j,k})$ at location
$x_{i,j,k}$, and a related magnetic energy $E_B = B^2 / (8 \pi)$ can be
defined. The instability criterion $\Delta B > z_{crit}$ involves than 
a magnetic gradient or magnetic field curvature (Charbonneau et al.~2001),
$$
	\Delta B = B_{i,j,k} - {1 \over 2S} \sum_{nn=1}^{2S} B_{nn} \ ,
	\qquad |\Delta B| > z_{crit}
        \eqno(2.2)
$$
where the summation includes all $2 \times S$ nearest neighbors 
($``nn''$) in a Cartesian S-dimensional lattice. 

More realistically, physics-based models have been attempted by applying the 
magneto-hydrodynamic (MHD) equations to the lattice field $B({\bf x})$,
obeying Amp\`ere's law for the current density ${\bf j}$,
$$
	{\bf j} = {1 \over 4\pi} ( \nabla \times {\bf B} ) \ ,	
        \eqno(2.3)
$$
which yields together with Ohm's law the induction equation,
$$
	{\partial {\bf B} \over \partial t}
	= \nabla \times ( \bf v \times {\bf B}) + \eta \nabla^2 {\bf B} \ ,
        \eqno(2.4)
$$
and fulfills the divergence-free condition for the magnetic field,
$$
	\nabla \cdot {\bf B} = 0 \ .
        \eqno(2.5)
$$
The instability threshold can then be expressed in terms of a critical
resisitivity $\eta$. Such cellular models with discretized 
magneto-hydrodynamics (MHD) have been applied to magnetospheric phenomena,
triggering magnetospheric substorms by perturbations in the solar wind 
(Takalo et al.~1993; 1999), as well as to solar flares 
(Vassiliadis et al.~1998; Isliker et al.~1998).
Initially, solar flares were modeled with isotropic magnetic field cellular 
automaton models (Lu and Hamilton 1991). However, since the plasma-$\beta$
parameter is generally smaller than unity in the solar corona, particle
and plasma flows are guided by the magnetic field, and thus SOC avalanches
are more suitably represented with anisotropic 1D transport 
(Vlahos et al.~1995). Other incarnations of cellular SOC models
mimic the magnetic field braiding that is thought to contribute 
to coronal heating (Morales and Charbonneau 2008).
On the other side, SOC models that involve discretized ideal MHD equations
have been criticized to be inadequate to describe the highly resistive
and turbulent evolution of magnetic reconnection processes, which are
believed to be the driver of solar flares.

Besides the magnetic field approach of the a physical variable in
SOC lattice simulations, which predicts the magnetic energy
$E_B \propto B^2 / (8 \pi)$ or current ${\bf j} \propto \nabla \times {\bf B}$
in each node point, we still have to model the physical units of observables. 
The magnetic field or electric currents can only be observed by in-situ
measurements, which is possible for magnetospheric or heliospheric
phenomena with spacecraft, but often observables can only be obtained
by remote-sensing measurements in form of photon fluxes emitted in various
wavelength ranges, such as for solar or astrophysical sources.
This requires physical modeling of the radiation process, a component
that has been neglected in most previous literature on SOC phenomena.

If the physical quantity $z_{i,j,k}$ at each cellular node is defined
in terms of an energy $e$, the discretized change of the quantity $dz/dt$ 
corresponds then to a quantized amount of dissipated energy $<\Delta e>$ 
per node. The instantaneous energy dissipation rate $de/dt =
<\Delta e> dV/dt$ of the system scales then with the instantaneous 
volume change $dV(t)/dt$, while the total amount of dissipated energy 
during an avalanche event can be computed from the time integral,
$E=<\Delta e> \int_0^t V(t) dt$, assuming that a mean energy quantum 
$<\Delta e>$ is dissipated per unstable node. 

In astrophysical applications, the observed quantity is usually a 
photon flux, and thus the energy quantity $e=h \nu$ corresponds to 
the photon energy at frequency $\nu$, while $<\Delta e>= N h \nu/\Delta t$ 
is the photon flux of $N$ photons that are radiated in a volume cell 
$\Delta V$ during a time step $\Delta t$ for a given radiation process.  
The conversion of the intrinsic energy dissipation rate $de/dt
= <\Delta e> (dV/dt)$ (e.g., of magnetic energy $E_B$ in a magnetic 
reconnection process) has then to be related to the amount of emitted 
photons $<\Delta e> =N h \nu/\Delta t$ by a physical model of the 
dissipation process. 
The physical model of the radiation process may involve a nonlinear
scaling between the amount of dissipated energy and the number of
emitted photons in a specific wavelength range, which changes the 
powerlaw slope of the observed distributions of photon fluxes, compared
with the predictions of the generic quantity $z_{i,j,k}$ used in SOC 
cellular automaton simulations. 

\section{Analytical SOC Models} 

The complexity of the spatio-temporal pattern of a SOC avalanche is
produced by a simple mathematical re-distribution rule that defines
the next-neighbor interactions on a microscopic level, while the
observed structure manifests itself at the macroscopic level. While cellular
automaton models operate on the microscopic level of a lattice cell,
the complex macroscopic structure cannot be analytically derived from 
the microscopic states of the nonlinear system. In classical
thermodynamics, the macroscopic state of a gas, such as the velocity
distribution function of molecules can be derived from the (binomial)
probability distribution function of microscopic states. In contrast,
nonlinear dissipative systems in the state of self-organized criticality
seem to exhibit a higher level of complexity, so that their macroscopic
morphology cannot be analytically derived from the microscopic states.
Therefore, analytical models can describe the macroscopic SOC parameters
only by simplified approximations of the SOC dynamics, which are aimed to
be consistent with the observed (powerlaw-like) probability or occurrence 
frequency distributions. We will describe a few of such analytical models 
in the following, which make quantitative predictions of the powerlaw
slopes and scaling laws between SOC parameters. Such analytical models
can then be used for Monte-Carlo simulations of SOC avalanches and
be forward-fitted to the observed distributions of SOC parameters.

\subsection{Exponential-Growth SOC Model (EG-SOC)} 

SOC avalanches have always a growth phase with nonlinear characteristics,
similar to a multiplicative chain-reaction. Mathematically, the simplest
function that grows in a multiplicative way is the exponential function
with a positive growth rate. It is therefore not surprising that the
first analytical models of SOC processes were based on such an
exponential growth function (even when they were not called SOC models
at the pre-Bak time). The attractive feature for SOC applications is
the fact that the simple assumption of an exponential growth curve
combined with random durations automatically leads to a powerlaw
distribution function, as we will see in the following.

The earliest analytical models in terms of an exponential growth phase
with saturation after a random time interval go back to
Willis and Yule (1922) who applied it to geographical distributions of plants
and animals. Yule's model was applied to cosmic rays (Fermi 1949), 
to cosmic transients and solar flares \index{solar flares}
(Rosner and Vaiana 1978; Aschwanden et al.~1998),    
to the growth dynamics of the world-wide web \index{web access}
(Huberman and Adamic 1999), as well as to the distribution of
the sizes of incomes, cities, internet files, biological taxa, and
in gene family and protein family frequencies (Reed and Hughes 2002). 

\begin{figure}
\centerline{\includegraphics[width=1.00\textwidth]{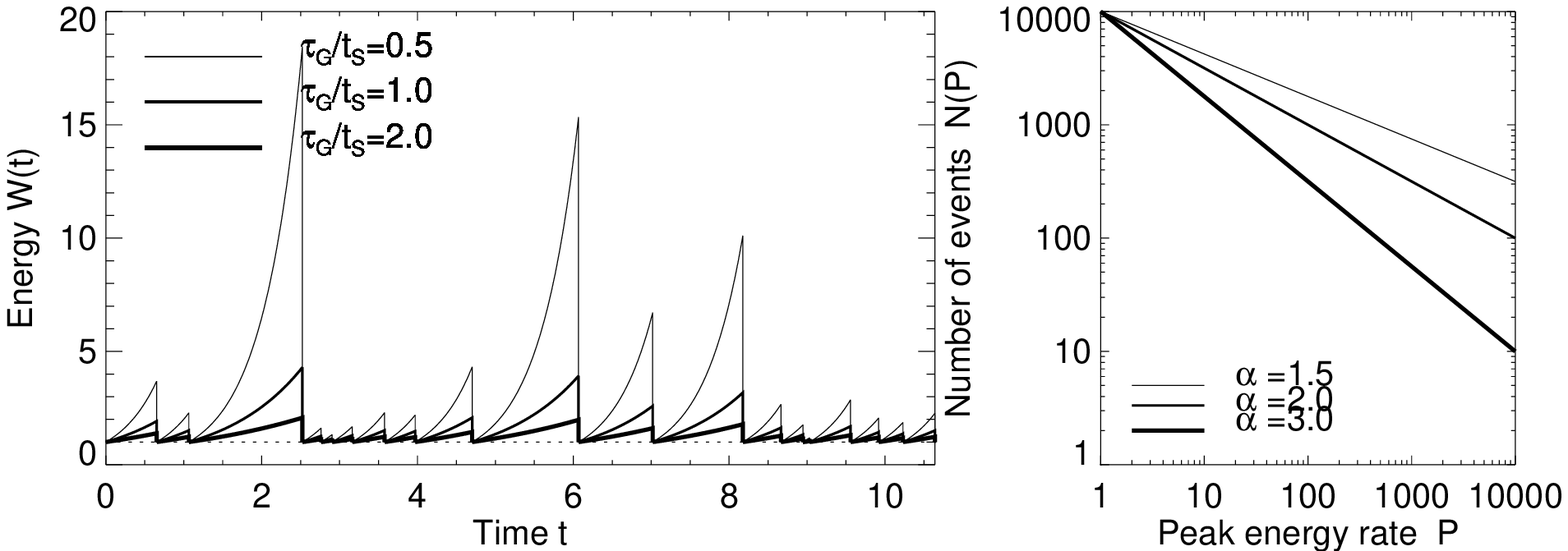}}
\captio{Time evolution of energy release rate $W(t)$ for 3
different ratios of growth times to saturation times,
$\tau_G/t_S=(0.5,1.0,2.0)$ (left) and the corresponding powerlaw
distributions of the peak energy release rate $P$. Note that the
event set with the shortest growth time ($\tau_G/t_S=0.5$)
reaches the highest energies and thus
produces the flattest powerlaw slope ($\alpha=1+\tau_G/t_S=1.5$).}
\end{figure}

Here we describe the analytical derivation of the exponential-growth
SOC model, following Aschwanden (2011a, Section 3.1).
We define the time evolution of the energy release rate $W(t)$ of a 
nonlinear process that starts at a threshold energy of $W_0$ by
$$
        W(t) = W_0 \ \exp{\left({t \over {\tau}_G} \right)} \ ,
        \qquad 0 \le t \le \tau \ ,
        \eqno(2.6)
$$
where ${\tau}_G$ represents the exponential growth time. The process
grows exponentially until it saturates at time $t=\tau$ with a
saturation energy $W_S$,
$$
        W_S = W(t=\tau) = W_0 \ \exp{\left({\tau \over {\tau}_G} \right)} \ .
        \eqno(2.7)
$$
We define a peak energy release rate $P$ that represents the maximum
energy release rate $W_S$, after subtraction of the threshold energy $W_0$,
that corresponds to the steady-state energy level before the nonlinear
growth phase,
$$
        P = W_S - W_0 = W_0 \left[ \exp{ \left( {\tau \over \tau_G} \right)}
                - 1 \right] \ .
        \eqno(2.8)
$$
In the following, we will refer to the peak energy release rate $P$
also briefly as ``peak energy''. For the saturation times $\tau$,
which we also call ``rise times'', we assume a random probability distribution,
approximated by an exponential function $N(\tau)$ with e-folding time constant
$t_S$,
$$
        N(\tau ) d\tau = {N_0 \over t_S}
        \exp{\left(-{\tau \over t_S}\right)} d\tau \ .
        \eqno(2.9)
$$
This probability distribution is normalized to the total number of $N_0$
events.

In order to derive the probability distribution $N(P)$ of peak energy
release rates $P$, we have to substitute the variable of the peak energy,
$P$, into the function of the rise time $\tau(P)$, which yields
(using the functional relationship $\tau(P)$ from Eq.~2.8),
$$
        N(P) dP = N(\tau ) d\tau =
        N[\tau(P)] \left| {d\tau \over dP} \right| dP 
                = {N_0 (\alpha_P - 1) \over W_0}
        \left({P \over W_0} + 1 \right)^{-\alpha_P} \ dP \ ,
        \eqno(2.10)
$$
which is an exact powerlaw distribution for large peak energies
($P \gg W_0$) with a powerlaw slope $\alpha_P$ of
$$
        \alpha_P = \left( 1 + {\tau_G \over t_S} \right) \ .
        \eqno(2.11)
$$
The powerlaw slope thus depends on the ratio of the growth time to the
e-folding saturation time, which is essentially the average number of
growth times. Examples of time series with avalanches of different
growth times ($\tau_G/t_S=0.5, 1.0, 2.0$) are shown in Fig.~2.3,
along with the corresponding powerlaw distributions of peak energies $P$.
Note that the fastest growing events produce the flattest powerlaw
distribution of peak energies.

\begin{figure}
\centerline{\includegraphics[width=0.70\textwidth]{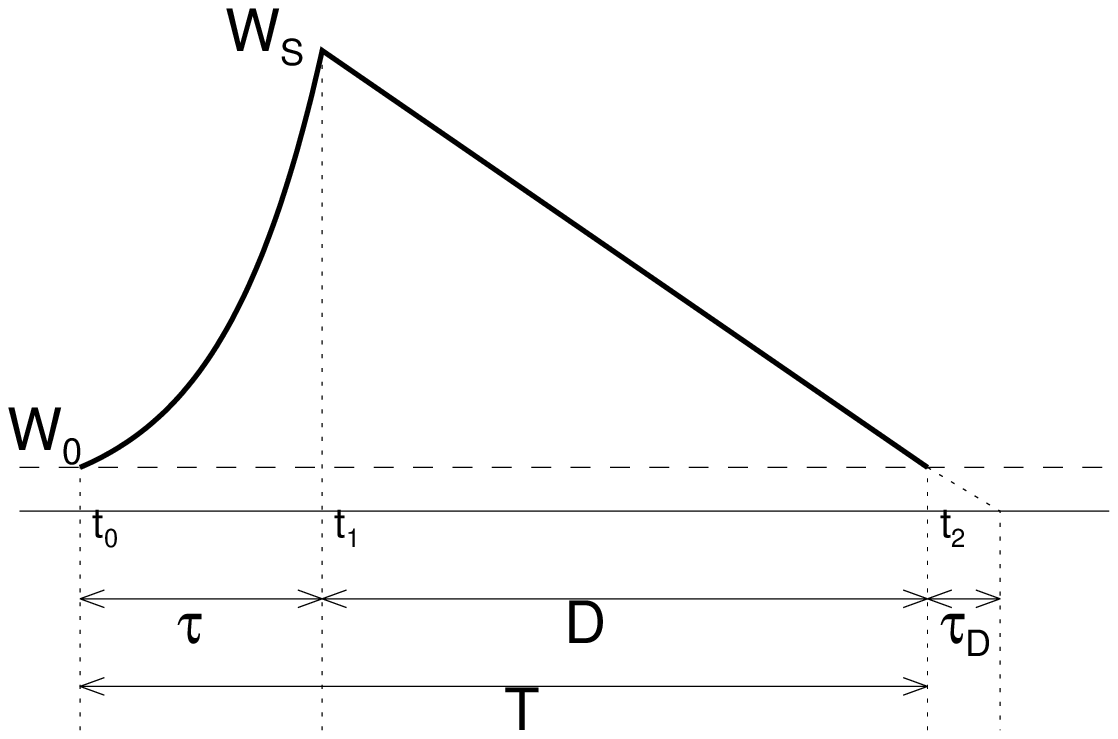}}
\captio{Schematic of the time evolution of an avalanche event, consisting
of (i) a rise time ($\tau$) with exponential growth of the energy
release $W(t)$ from a threshold level $W_0$ to the saturation level $W_S$,
and (ii) a decay time ($D$) with a constant decay rate
$\eta = dW/dt = W_0/\tau_D$.}
\end{figure}

Once an instability has released a maximum amount $W_S$ of energy,
say when an avalanche reaches its largest velocity on a sandpile,
the energy release gradually slows down until the avalanche comes
to rest. For sake of simplicity we assume a linear decay phase 
of the released energy (Fig.~2.4),
$$
        W(t) = W_0 + (W_S-W_0)
        \left( 1 - {(t - t_1) \over D} \right) \quad
        t_1 < t < t_2 \ ,
        \eqno(2.12)
$$
where $t_2$ is the end time of the process at $t_2 = t_1 + D$.
The time interval $T$ of the total duration of the
avalanche process is then the sum of the exponential rise phase $\tau$
and the linear decay phase $D$ as illustrated in Fig.~2.4,
$$
        T = \tau + D = \tau_G \ln{\left({P \over W_0} + 1 \right)}
                     + \tau_D {P \over W_0} \ .
        \eqno(2.13)
$$
We see that this relationship predicts an approximate proportionality of
$T \propto P$ for large avalanches, since the second term, which is
linear to P, becomes far greater than the first term with a
logarithmic dependence ($\propto \ln{P}$).

For the calculation of the distribution $N(\tau)$ we express
the total duration $T$ in terms of the rise time $\tau$
and find a powerlaw function for the distribution of flare durations $T$,
$$
        N(T) dT = N[\tau(T)] \left| {d\tau \over dT} \right| dT \\
        = {N_0 (\alpha_T - 1) \over \tau_D}
        \left( {T \over \tau_D} + 1 \right)^{-\alpha_T} \ .
        \eqno(2.14)
$$
We define also the total released energy $E$ by the time integral of the
energy release rate $W(t)$ during the event duration $T$, but neglect the
rise time $\tau$ (i.e., $T \approx D$) and subtract the
threshold level $W_0$ before the avalanche,
$$
        E = \int_0^{T} \left[ W(t)-W_0 \right] \ dt
        \approx \int_{\tau}^{\tau+D} \left[ W(t)-W_0 \right] \ dt
        = {1\over 2} P D      \ .
        \eqno(2.15)
$$
leading to a powerlaw-like function for the frequency distribution 
of energies $E$,
$$
        N(E) dE = N[P(E)] \left| {dP \over dE} \right| dE \\
        = {N_0 (\alpha_P - 1) \over 2 E_0}
        \left[ \sqrt{ {E \over E_0} } + 1 \right]^{-\alpha_P}
        \left[ {E \over E_0} \right]^{-1/2}
        \eqno(2.16)
$$
Thus, we find the following approximate scaling laws between the
powerlaw indices,
$$
        \begin{array}{ll}
        \alpha_P = 1 + {\tau_G / t_S} \\
        \alpha_T = \alpha_P       \\
        \alpha_E = (\alpha_P+1)/2 \\
        \end{array}
        \eqno(2.17)
$$

\begin{figure}
\centerline{\includegraphics[width=1.0\textwidth]{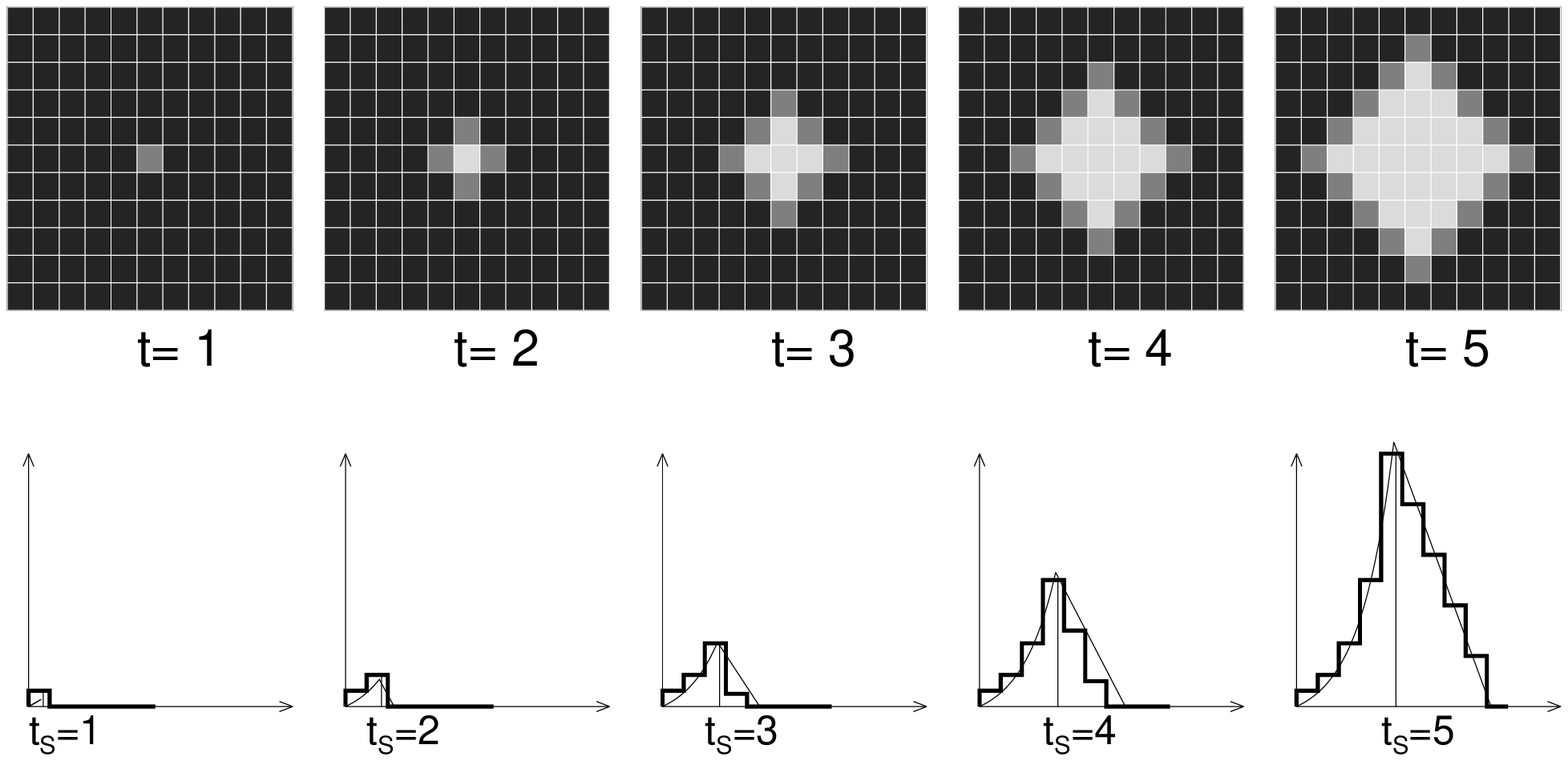}}
\captio{Spatial patterns of a propagating avalanche in subsequent
time steps in a 2-D cellular automaton model with a next-neighbor
redistribution rule (top) and time profiles of energy release rate
(bottom), for saturation times of $t_S=1,2,...,5 \Delta t$.
The black cells represent
cells with random fluctuations below the threshold, $z_k<z_c$,
the grey cells contain possibly unstable cells with fluctuation
$z_k \ge z_c$ that are subject to a first redistribution, while
the white cells have already been affected by a redistribution rule
before. Most avalanches die out after step t$\gapprox 2$.}
\end{figure}

In summary, this model predicts powerlaw distribution functions for
the three SOC parameters $P$, $E$, and $T$, which match the simulated
distributions with cellular automaton simulations, as well as the
observed distributions of solar flare hard X-ray fluxes (Lu and Hamilton
1991). For a ratio of ${\tau_G/t_S}=1$, this model predicts
$\alpha_P=2.0$, $\alpha_T=2.0$, and $\alpha_E=1.5$. This particular
time ratio of ${\tau_G/t_S}=1$ implies that an avalanche typically
saturates after one exponential growth time (see the spatio-temporal
patterns for small avalanches in Fig.~2.5). Open 
questions are: Which physical process would explain this particular time 
scale ratio $\tau_G/t_S=1$ ? Why does the decay phase has a linear behavior? 
Is the exponential distribution of avalanche growth times consistent
with observations, since powerlaw-like distributions are expected
for SOC parameters? How can the observed intermittent fluctuations
of time profiles and the fractal geometry be accomodated in a model with 
a monotonic growth function? Although this model seems not to reproduce
all observed properties of SOC phenomena, it has a didactical value,
since it represents the most basic model that links the nonlinear 
evolution of instabilities to the powerlaw distributions observed
in SOC phenomena.

\medskip
The exponential-growth model is most suitable for 
multiplicative avalanche processes, where the increase
per time step during the rise phase is based on a multiplicative factor,
such as it occurs in nuclear chain reactions, population growth, or
urban growth. Alternatively, avalanche processes that continuously 
expand in space may show an energy increase that scales with the area or 
volume, which has a powerlaw relationship to the spatial or temporal
scale, i.e., $A(t) \propto r(t)^2 \propto t^2$ or
$V(t) \propto r(t)^3 \propto t^3$. Such a model with a powerlaw-growth
function
$$
        W(t) = W_0 \ \left[ 1 + \left({t \over \tau_G}\right)^p \right] \ ,
        \eqno(2.18)
$$
rather than an exponential-growth function (Eq.~2.6) was computed in
Aschwanden (2011a; Section 3.2), which predicts identical scaling laws
between the SOC parameters ($P, E, T$), but the occurrence frequency 
distributions exhibit an exponential fall-off at the upper end.
Otherwise it has similar caveats as the exponential-growth model
(i.e., monotonic growth curve rather than intermittency, and 
Euclidean rather than fractal avalanche volume).

\medskip
Another variant of the exponential-growth model is the
logistic-growth model (Aschwanden 2011a; Section 3.3), which has a
smoother transition from the initially exponential growth to the
saturation phase, following the so-called {\sl logistic equation},
$$
        {dE(t) \over dt } = {\ E(t) \over \tau_G } \cdot
        \left[1 - {E(t) \over E_\infty} \right] \ ,
        \eqno(2.19)
$$
where the dissipated energy is limited by the so-called {\sl carrying
capacity} limit $E_{\infty}$ used in ecological applications. The
resulting occurrence frequency distributions are powerlaws for
the energy $E$ and peak energy dissipation rate $P$, but exponential
functions for the rise time $\tau$ and duration $T$ of an avalanche.
Like the exponential-growth and the powerlaw-growth model, the temporal
intermittency and the geometric fractality observed in real SOC
phenomena are not reproduced by the smoothly-varying time evolution
of the logistic-growth model.

\subsection{The Fractal-Diffusive SOC Model (FD-SOC)}

The microscopic structure of a SOC avalanche has been simulated with
a discretized mathematical re-distribution rule, which leads to highly
inhomogeneous, filamentary, and fragmented topologies during the
evolution of an avalanche. Therefore, it appears to be adequate to
develop an analytical SOC model that approximates the inhomogeneous
topology of an avalanche with a fractal geometry. Bak and Chen (1989)
wrote a paper entitled ``The physics of fractals'', which is
summarized in their abstract with a single sentence: {\sl Fractals
in nature originate from self-organized critical dynamical processes}.

A statistical fractal-diffusive avalanche model of a slowly-driven
SOC system has been derived in Aschwanden (2012). This analytical model 
represents a universal (physics-free) description of the statistical 
time evolution and occurrence frequency distribution function of SOC 
processes. It is based on four fundamental assumptions: 
(1) A SOC avalanche grows spatially like a
diffusive process; (2) The spatial volume of the instantaneous energy
dissipation rate is fractal; (3) The time-averaged fractal dimension
is the mean of the minimum dimension $D_{S,min}\approx 1$ (for a sparse
SOC avalanche) and the maximum dimension $D_{S,max}=S$ (given by the Euclidean
space); and (4) The occurrence frequency distribution of length scales
is reciprocal to the size $L$ of spatial scales, i.e., $N(L) \propto L^{-S}$
in Euclidean space with dimension $S$. We will discuss these assumptions
in more detail in the following.

The first assumption of a diffusive process is based on numerical
simulations of cellular automaton models. A SOC avalanche propagates
in a cellular automaton model by next-neighbor interactions in a critical
state, where energy dissipation propagates only to the next neighbor cells
(in a S-dimensional lattice grid) that are above a critical threshold.
This mathematical rule that describes the entire dynamics and evolution
of a SOC avalanche is very simple for a single time step, but leads to
extremely complex spatial patterns after a finite number of time steps.
For a visualization of a large number of such complex spatial patterns
generated by a simple iterative mathematical redistribution rule see,
for instance, the book ``A New Kind of Science'' by Wolfram (2002).
The complexity of these spatial patterns can fortunately be characterized
with a single number, the fractal dimension $D_S$. If one monitors the
time evolution of a spatial pattern of a SOC avalanche in a cellular
automaton model, one finds that the length scale $x(t)$ evolves with time
approximately with a diffusive scaling (see radius $r(t)$ of snapshots
of a 2-D cellular automaton evolution in Fig.~(2.6) and its time evolution 
$r(t) \propto t^{1/2}$ in Fig.~(2.8), bottom right panel),
$$
        x(t) \propto t^{1/2} \ ,
        \eqno(2.20)
$$
which leads to a statistical scaling law between the avalanche sizes 
$L=x(t=T)$ and time durations $T$ of SOC avalanches,
$$
        L \propto T^{1/2} \ .
        \eqno(2.21)
$$

\begin{figure}
\centerline{\includegraphics[width=1.0\textwidth]{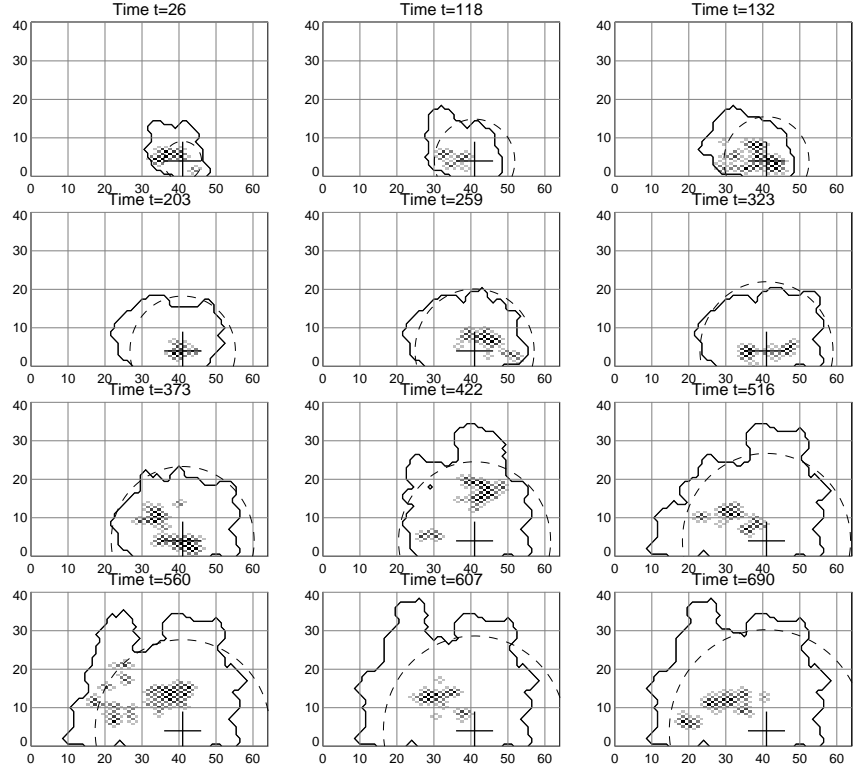}}
\caption{Time evolution of a large avalanche event in
a 2-D cellular automaton simulation with grid size $N=64^2$.
The 12 panels show snapshots at particular burst times from
$t=26$ to $t=690$ when the energy dissipation rate peaked.
Active nodes where energy dissipation occurs at time $t$ are visualized
with black and grey points, depending on the energy dissipation level.
The starting point of the avalanche occurred at pixel $(x,y)=(41,4)$,
which is marked with a cross. The time-integrated envelop of the
avalanche is indicated with a solid contour, and the diffusive
avalanche radius $r(t) = t^{1/2}$ is indicated with a dashed circle
(Aschwanden 2012).}
\end{figure}

\begin{figure}
\centerline{\includegraphics[width=1.0\textwidth]{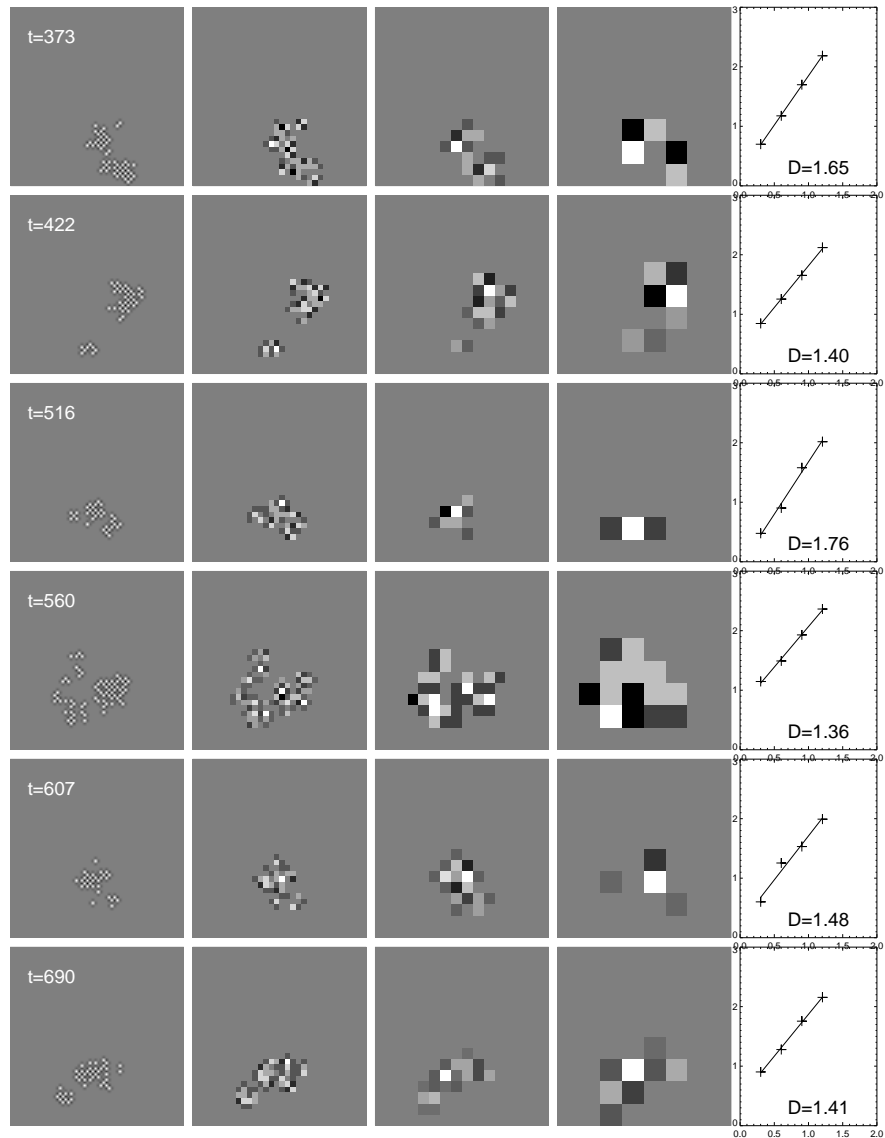}}
\caption{Determination of the fractal dimension $D_2=\log{A_i}/\log{x_i}$
for the instantaneous avalanche sizes of 6 time steps of the
avalanche event shown in Fig.~(2.6). Each row is a different time step
and each column represents a different binning of macropixels
($\Delta x_i=1,2,4,8$). The fractal dimension is determined by a linear
regression fit shown on the right-hand side. The mean fractal
dimension of the 12 avalanche snapshots shown in Fig.~(2.6) is 
$D_2=1.43\pm0.17$ (Aschwanden 2012).}
\end{figure}

\begin{figure}
\centerline{\includegraphics[width=1.0\textwidth]{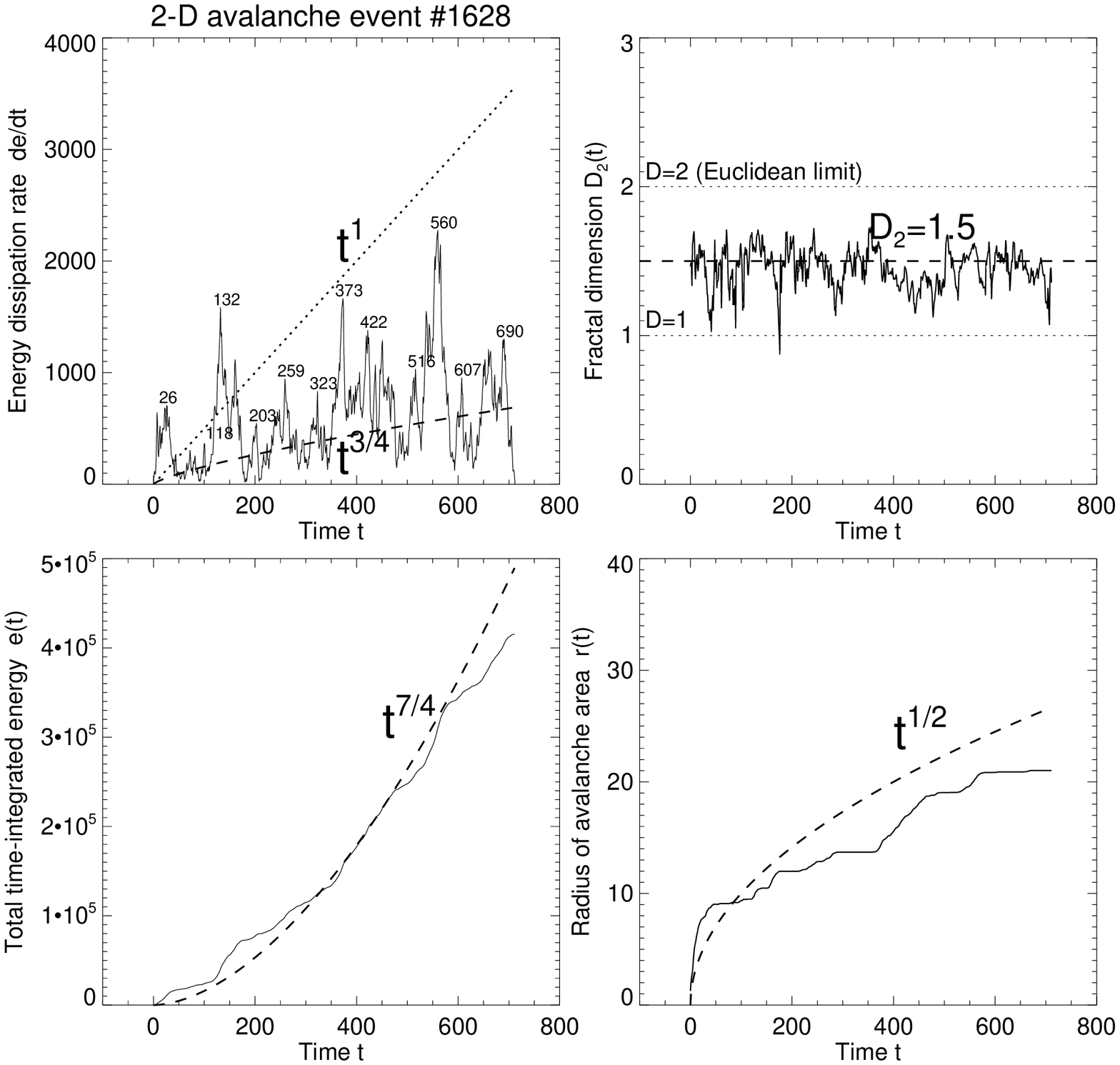}}
\caption{Time evolution of the same large avalanche event from a 2-D 
cellular automaton simulation with grid size $N=64^2$ as shown in Fig.~(2.6).
The time profiles include the instantaneous energy dissipation rate
$f(t)=de/dt$ (top left),
the time-integrated total energy $e(t)$ (bottom left),
the instantaneous fractal dimension $D_2(t)$  (top right),
and the radius of the avalanche area $r(t)$ (bottom right).
The observed time profiles from the simulations are outlined in
solid linestyle and the theoretically predicted average evolution
in dashed linestyle. The statistically predicted values of the
instantaneous energy dissipation rate $f(t) \propto t^{3/4}$ (dotted curve)
and peak energy dissipation rate $p(t) \propto t^1$ (dashed curve)
after a time interval $t$ are also shown (top left panel).
The 12 time labels from 26 to 690 (top left frame) correspond
to the snapshot times shown in Fig.~(2.6) (Aschwanden 2012).}
\end{figure}

The second assumption of a fractal pattern of the instantaneous energy
rate is also based on tests with cellular automaton simulations 
(see measured fractal dimensions $D_2$ of snapshots in Fig.~(2.7) 
and the time evolution $D_2(t)$ in Fig.~(2.8) top right panel). 
The fractal dimension is essentially a
simplified parameter that describes the ``micro-roughness'', ``graininess'',
or inhomogeneity of critical nodes in a lattice grid in the state of
self-organized criticality. Of course, such a single number is a gross
over-simplification of a complex system with a large number of degrees of
freedom, but the numerical simulations confirm that avalanche patterns
are fractal (Fig.~2.7). Thus we define the volume $V_S(t)$ of 
the instantaneous energy dissipation rate in terms of a fractal
(Hausdorff) dimension $D_S$ that scales with the length scale $x$ as,
$$
        V_S(t) \propto x^{D_S} \ ,
        \eqno(2.22)
$$
which leads also to a statistical scaling law between avalanche
volumes $V$ and spatial scales $L$ or durations $T$ of SOC avalanches
(with Eq.~2.21) 
$$
        V_S \propto L^{D_S} \propto T^{D_S/2} \ .
        \eqno(2.23)
$$

The third assumption of the mean fractal dimension has also been
confirmed by numerical simulations of cellular automaton SOC processes
in all three dimensions $S=1,2,3$ (Aschwanden 2012), but it can also
be understood by the
following plausibility argument. The sparsest SOC avalanche that
propagates by next-neighbor interactions is the one that spreads only
in one spatial dimension, and thus yields an estimate of the minimum
fractal dimension of $D_{S,min} \approx 1$, while the largest SOC avalanche
is almost space-filling and has a volume that scales with the Euclidean
dimension, $D_{S,max}=S$. Combining these two extremal values, we
can estimate a time-averaged fractal dimension $<D_S>$ from the 
arithmetic mean,
$$
        <D_S> \approx {D_{S,min}+D_{S,max} \over 2} = {1 + S \over 2} \ ,
        \eqno(2.24)
$$
which yields a mean fractal dimension of $<D_3>=(1+3)/2=2.0$ for the
3D case $(S=3)$.

\begin{figure}
\centerline{\includegraphics[width=1.0\textwidth]{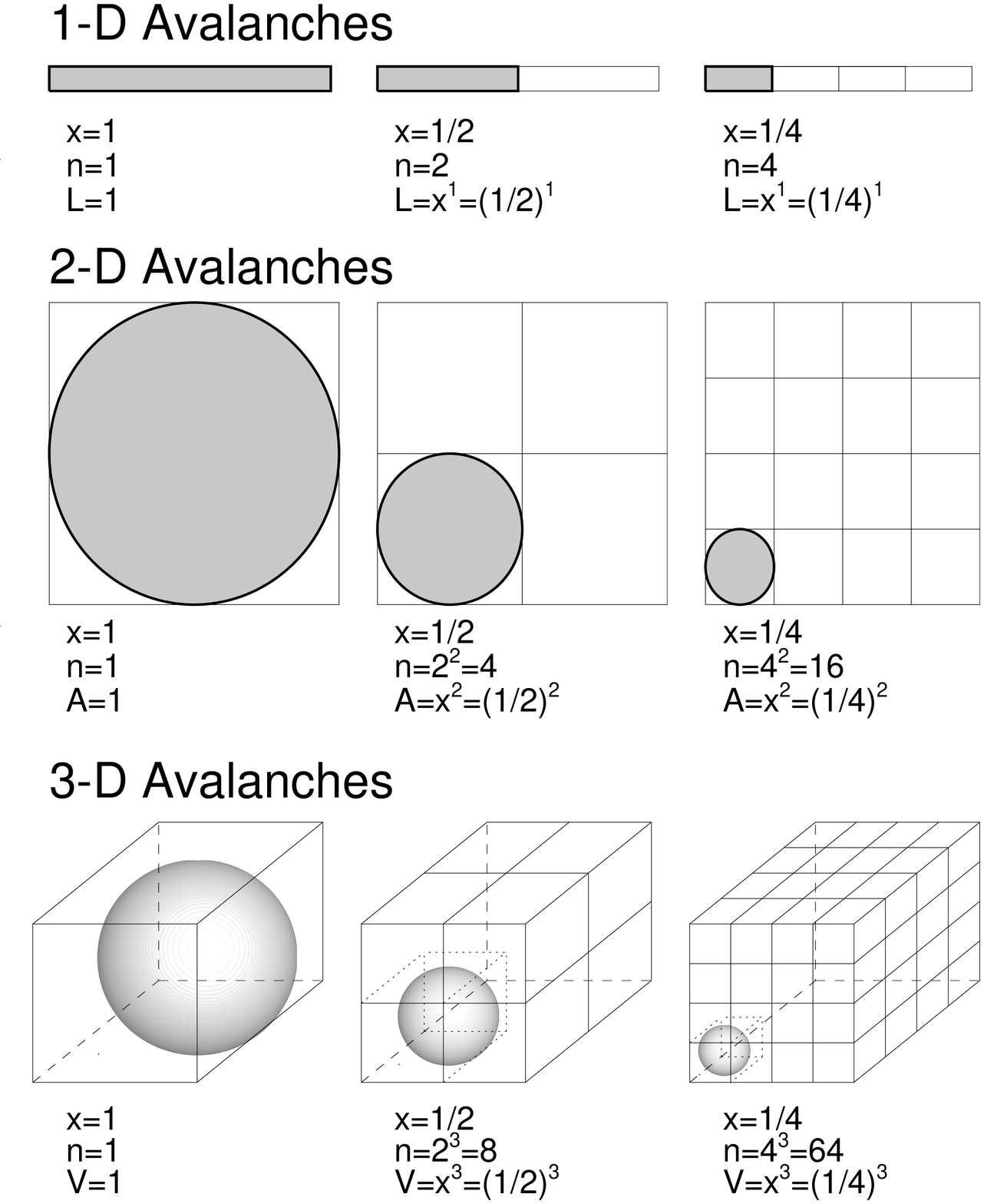}}
\caption{Schematic diagram of the Euclidean volume scaling
of the diffusive avalanche boundaries, visualized as circles or spheres
in the three Euclidean space dimensions $S=1,2,3$. The Euclidean
length scale $x$ of subcubes decreases by a factor 2 in each step
($x_i=2^{-i}, i=0,1,2$), while the number of subcubes increases by
$n_i=(2^i)^S$, defining a probability of $N(x_i) \propto x_i^{-S}$
for each avalanche size with size $x_i$.}
\end{figure}

The fourth assumption on the size distribution is a probability argument.
The system size $L_{sys}$ of a SOC system represents an upper limit of
spatial scales $L$ for SOC avalanches, i.e., $L \le L_{sys}$. For the
3D-case, the volumes $V$ of individual avalanches are also bound by
the volume $V_{sys}$ of the system size, i.e., $V = L^3 \le V_{sys} =
L_{sys}^3$. If the entire system is in a critical state, SOC avalanches
can be produced everywhere in the system, and the probability $N(L)$ for a
fixed avalanche size L with volume V is simply reciprocal to the size,
(visualized in Fig.~2.9), i.e.,
$$
        N(V) \propto {V_{sys} \over V} \propto V^{-1} \ ,
        \eqno(2.25)
$$
which is equivalent to
$$
        N(L) \propto {L_{sys}^3 \over L^3} \propto L^{-3} \ .
        \eqno(2.26)
$$

Based on these four model assumptions we can now quantify the time
evolution of SOC parameters. For sake of simplicity we apply this
fractal-diffusive SOC model to an astrophysical source that emits
a photon flux $f(t)$ that is proportional to the instantaneous
energy dissipation volume $V_S(t)$ of a SOC avalanche, where
a mean energy quantum $<\Delta E>$ is emitted per volume cell
element $\Delta V$. Thus, the emitted flux is (using the
diffusive scaling $x(t) \propto t^{1/2}$),
$$
        f(t) ={de(t) \over dt} \propto <\Delta E> V_S(t)
        =<\Delta E> x(t)^{D_S}
        =<\Delta E> t^{D_S/2} \ .
        \eqno(2.27)
$$
In the 2-D case with $D_S=D_2=(1+2)/2=3/2$ (Eq.~2.24) we expect a
statistical scaling of $f(t) \propto t^{3/4}$ (see Fig.~2.8 top left).
In the 3-D case with $D_S=D_3=(1+3)/2=2$, we expect than the proportionality
$f(t) \propto t^1$. In Fig.~(2.10) (second panel) we simulate such a 
flux time profile $f(t)$ by applying noise fluctuations in the time evolution 
of the fractal dimension $D_2(t)$ (Fig.~2.10, top).

The statistical peak value $p(t)$ of the energy dissipation rate
after time $t$ can be estimated from the largest possible avalanches,
which have an almost space-filling dimension $D_S \lapprox S$, 
$$
        p(t) ={de(t) \over dt} \propto <\Delta E> V_{S,max}(t)
        =<\Delta E> x(t)^S
        =<\Delta E> t^{S/2} \ .
        \eqno(2.28)
$$
and thus would be expected to scale as $p(t) \propto t^{S/2} \propto t^{1.5}$
for the $S=3$ case. Thus the time profile of peak values envelopes the
maximum fluctuations of the flux time profile $f(t)$ (Fig.~2.10, second panel).

The evolution of the total dissipated energy $e(t)$ after time $t$
is simply the time integral, for which we expect
$$
        e(t) = \int_0^t {de(\tau) \over d\tau} d\tau
        \propto \int_0^t \tau^{D_S/2} \propto t^{(1+D_S/2)} \ ,
        \eqno(2.29)
$$
which yields the function $e(t) \propto t^{7/4}$ for the 2-D case
(Fig.~2.8, bottom left), and a function $e(t) \propto t^2$ for the 3-D case
(Fig.~2.10, third panel). These time evolutions apply to every 
SOC model that has an emission $f(t)$ proportional to the
fractal avalanche volume $V_S(t)$. For applications to observations
in a particular wavelength range there may be an additional 
scaling law between the avalanche volume and emission (or intensity)
of the underlying radiation mechanism. 

\medskip
The fourth assumption on the probability distribution of (avalanche) 
length scales,
$$
       N(L) \propto V_S^{-1} \propto L^{-S} \ .
	\eqno(2.30)
$$
is a simple probability argument, implying that the number or occurrence 
frequency of avalanches has an equal likelyhood throughout the system, so it 
assumes a homogeneous distribution of critical states across the entire 
system. This is an extremely important assumption, which automatically
predicts a powerlaw distribution for length scales. Using the scaling
laws that result from Eq.~(2.21) and (2.27)-(2.29) for $L=x(t=T)$,
$E=e(t=T)$, $F=f(t=T)$, and $P=p(t=T)$ for an avalanch duration time $T$,
$$
        \begin{array}{l}
		L \propto T^{1/2}    \\
                F \propto T^{D_S/2}  \\
                P \propto T^{S/2}    \\
                E \propto T^{1+D_S/2}\\
        \end{array} \ ,
	\eqno(2.31)
$$
we can directly calculate the occurrence frequency distributions
for all these SOC parameters, by substituting the variables from the
correlative relationships given in Eq.~(2.31), yielding 
$$
        N(T) dT = N(L[T]) \left| {dL \over dT} \right| dT
        \propto T^{-[(1+S)/2]} \ dT \ .
	\eqno(2.32)
$$
$$
        N(F) dF = N(T[F]) \left| {dT \over dF} \right| dF
        \propto F^{-[1+(S-1)/D_S]} \ dF \ ,
	\eqno(2.33)
$$
$$
        N(P) dP = N(T[P]) \left| {dT \over dP} \right| dP
        \propto P^{-[2-1/S]}
        \ dP \ ,
	\eqno(2.34)
$$
$$
        N(E) dE = N(T[E]) \left| {dT \over dE} \right| dE
        \propto E^{-[1+(S-1)/(D_S+2)]}
        \ dE \ .
	\eqno(2.35)
$$
This derivation yields naturally powerlaw functions for all
parameters $L$, $T$, $F$, $P$, and $E$, which are the hallmarks of SOC systems.
In summary, if we denote the occurrence frequency distribution
$N(x)$ of a parameter $x$ with a powerlaw distribution with power index
$\alpha_x$,
$$
        N(x) dx \propto x^{-\alpha_x} \ dx \ ,
	\eqno(2.36)
$$
we have the following powerlaw coefficients $\alpha_x$ for the parameters
$x=T, F, P$, and $E$,
$$
        \begin{array}{ll}
        \alpha_T &= (1+S)/2 \\
        \alpha_F &=  1+(S-1)/D_S \\
        \alpha_P &=  2-1/S \\
        \alpha_E &=  1+(S-1)/(D_S+2)\\
        \end{array} \ .
	\eqno(2.37)
$$
The powerlaw coefficients $\alpha_x$ and correlation are summarized in Table 
2.2 separately for each Euclidean dimension $S=1,2,3$.

\begin{table}
\begin{center}
\normalsize
\caption{Theoretically predicted occurrence frequency
distribution powerlaw slopes $\alpha$ and power indices $\beta$
of parameter correlations predicted for SOC cellular automatons
with Euclidean space dimensions $S=1,2,3$, for the fractal dimension
$D_S$, length scale $L$, time duration $T$, instantaneous energy
dissipation rate (or flux) $F$, peak energy dissipation rate 
(or peak flux) $P$, and total time-integrated energy $E$.}
\medskip
\begin{tabular}{|l|l|l|l|}
\hline
Theory          &  S=1  &  S=2  &  S=3  \\
\hline
\hline
$D_S=(1+S)/2$ 	&1      &3/2    &2\\
$\alpha_L=S$ 	&1      &2      &3\\
$\alpha_T=(1+S)/2$ &1      &3/2    &2\\
$\alpha_F=1+(S-1)/D_S$ &1 &5/3    &2\\
$\alpha_P=2-1/S$  &1      &3/2    &5/3 \\
$\alpha_E=1+(S-1)/(D_S+2)$ &1 &9/7 &3/2 \\
\hline
$L \propto T^{1/2}$   &$L\propto T^{1/2}$  &$L\propto T^{1/2}$ &$L\propto T^{1/2}$\\
$F \propto T^{D_S/2}$ &$F\propto T^{1/2}$  &$F\propto T^{3/4}$ &$F\propto T^{1}  $\\
$P \propto T^{S/2}$   &$P\propto T^{1/2}$  &$P\propto T^{1}$   &$P\propto T^{3/2}$\\
$E \propto T^{1+D_S/2}$&$E \propto T^{3/2}$&$E \propto T^{7/4}$&$E \propto T^2$  \\
\hline
\end{tabular}
\end{center}
\end{table}

These correlation coefficients and powerlaw indices of frequency
distributions have been found to agree within $\approx 10\%$ with
numerical simulations of cellular automatons for all three Euclidean
dimensions ($S=1,2,3$) (Aschwanden 2011a). Some deviations, especially
fall-offs at the upper end of powerlaw distributions, are likely
to be caused by finite-size effects of the lattice grid.

\subsection{Astrophysical Scaling Laws} 

SOC theory applied to astrophysical observations covers many
different wavelength regimes, for instance gamma-rays, hard X-rays,
soft X-rays, and extreme ultra-violet (EUV) in the case of solar flares. 
A comprehensive review of such studies is given in Section 7 of 
Aschwanden (2011a).
However, since each wavelength range represents a different physical
radiation mechanism, we have to combine now the physics of
the observables with the (physics-free) SOC statistics.

\begin{figure}
\centerline{\includegraphics[width=0.9\textwidth]{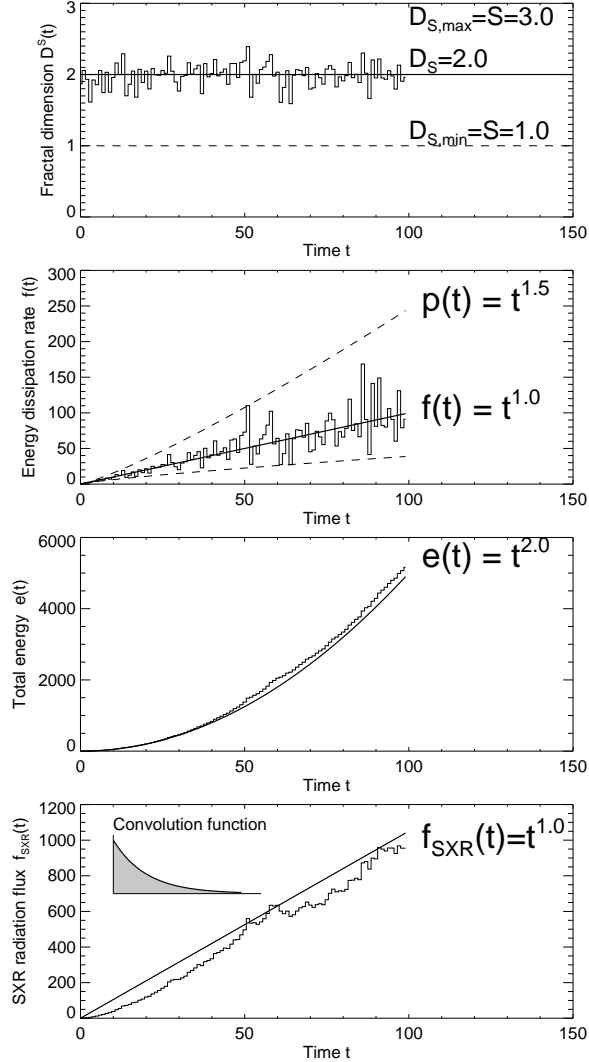}}
\caption{Simulation of the fractal-diffusive SOC model for an
Euclidean dimension $S=3$, showing the time evolution of the fractal
dimension $D_{S}(t)$ (top panel), the instantaneous energy dissipation
rate $f(t)$ and peak energy dissipation rate $p(t)$ (second panel),
the total time-integrated dissipated energy $e(t)$ (third panel),
and the soft X-ray
time profile $f_{sxr}(t)$ (bottom panel), which results from the
convolution of the instantaneous energy dissipation rate $f(t)$ (second
panel) with an exponential decay function with an e-folding time of
$\tau_{decay}$ (shown in insert of bottom panel) (Aschwanden and
Freeland 2012).}
\end{figure}

\medskip
Let us consider soft and hard X-ray emission in solar or stellar flares.
Soft X-ray emission during solar flares is generally believed to result
from thermal free-free and free-bound radiation of plasma that is
heated in the chromosphere by precipitation of non-thermal electrons
and ions, and which subsequently flows up into coronal flare (or
post-flare) loops, a process called ``chromospheric evaporation
process'' (for a review see, e.g., Aschwanden 2004).
Therefore, we can consider the flare-driven chromospheric
heating rate as the instantaneous energy dissipation process of a
SOC avalanche, as shown in the simulated function $f(t)$ in Fig.~2.10
(second panel). The heated plasma, while it fills the coronal flare
loops, loses energy by thermal conduction and by radiation of
soft X-ray and EUV photons, which generally can be characterized
by an exponential decay function after an impulsive heating spike.
In Fig.~2.10 (bottom) we mimic such a soft X-ray radiation light curve
by convolving the instantaneous energy dissipation rate $f(t)$
(Fig.~2.10, second panel) with an exponentially decaying radiation
function (with an e-folding time constant of $\tau_{decay})$,
$$
        f_{sxr}(t) = \int_{-\infty}^t f(t) 
	\exp{\left[-{(t-t') \over \tau_{decay}} \right]} dt' \ ,
	\eqno(2.38)
$$
which shows also a time dependence that follows approximately
$$
        f_{sxr}(t) \propto f(t) \propto t^{1.0} \ ,
	\eqno(2.39)
$$
because the convolution with an exponential function with a constant
e-folding time constant acts like a constant multiplier.
In the limit of infinitely long decay times $(\tau_{decay} \mapsto \infty)$,
our convolution function (Eq.~2.38) turns into a time integral of the
heating function $f(t)$, which is also known as Neupert effect
(Dennis and Zarro 1993; Dennis et al.~2003).

The heating function is identified with the non-thermal hard X-ray
emission, which indeed exhibits a highly fluctuating and intermittent
time profile for energies of $\gapprox 25$ keV, where non-thermal emission
dominates, 
$$
        f_{hxr}(t) \propto p(t) \propto t^{1.5} \ ,
	\eqno(2.40)
$$
Thus the occurrence frequency distributions of fluxes are expected to be
different for soft X-rays and hard X-rays, the hard X-ray flux $f_{hxr}(t)$
follows the statistics of the highly fluctuating peak energy dissipation
rate $p(t)$, while the soft X-ray flux $f_{sxr}(t)$ is expected to
follow the statistics of the smoothly-varying (convolved) time profile
$f(t)$. The total duration $T$ of energy release of an avalanche (or
flare here) corresponds essentially to the rise time $t_{rise}$ of 
the soft X-ray flux, or to the total flare duration for hard X-rays, 
because the decay phase of a soft X-ray flare light curve is dominated by
conductive and radiative loss, rather than by continued heating input.
Thus based on the generic relationships summarized in Table 2.2
we expect for the 3-D case ($S=3$),
$$
        N(T)       \propto T^{-\alpha_T} = T^{-2} \ ,
	\eqno(2.41)
$$
$$
        N(f_{sxr}) \propto F^{-\alpha_F} = F^{-[1+(S-1)/D_S]} = F^{-2} \ ,
	\eqno(2.42)
$$
$$
        N(f_{hxr}) \propto P^{-\alpha_P} = P^{-[2-1/S]} = P^{-5/3} \ .
	\eqno(2.43)
$$
Applications to observations did show satisfactory agreement with these
theoretical values, i.e. powerlaw slopes of $\alpha_F=2.0$ for
soft X-rays and $\alpha_P=1.67$ for hard X-rays (Aschwanden 2011a; 2011b;
Aschwanden and Freeland 2012), except for
the occurrence frequency distributions of flare durations $T$ during
solar cycle maxima, when the flare pile-up bias appears to have
a steepening side-effect (Aschwanden 2012).

\subsection{Earthquake Scaling Laws}

While the foregoing discussion is relevant to astrophysical SOC phenomena
(solar and stellar flares), similar physical scaling relationships between
the observables and SOC cellular automaton quantities $z_{i,j,k}$ can
be developed in other fields. For earthquakes in geophysics for instance,
measured quantities include the length $L_S$ and width $L_w$ of a
rupture area, so the rupture area has the scaling $A \propto L_S L_w$.
For large ruptures, however, when the surface rupture length is much
larger than the rupture width, i.e., $L_s \gg L_w$,
the width $L_w$ was found to be approximately constant, so that large
ruptures saturate the faults surface and scale with a fractal dimension
of $D=1$, whereas a smaller rupture propagates across both dimension
$(D=2)$ of the fault face (Yoder et al. 2012). The modeling of 
size distributions of earthquake magnitudes can then be conducted with
a similar methodology as outlined with our fractal-diffusive model
(Section 2.2.2), requiring: (i) The definition of the Euclidean space
dimension for earthquakes, which could be $S=2$ if they are treated as
surface phenomena without significant depth variability, or by $S=3$
otherwise; (ii) The measurement of the fractal dimension $D$, which
can exhibit multi-fractal scaling from $D=1$ for large earthquakes
to $D=2$ for small earthquakes; and (iii) assigning the observable
$m$, which is the magnitude of the earthquake, to the corresponding
physical quantity, i.e., the peak energy dissipation rate $P$, the
smoothed energy dissipation rate $F$, or total time-integrated energy $E$.
Furthermore, earthquake size distributions are often reported in terms of
a cumulative distribution function $N(>m)$, which has a powerlaw index
$\beta \approx  \alpha-1$ that is flatter by one than the differential
distribution function $N(m) dm$.

\section{Alternative Models Related to SOC} 

Here we discuss a number of alternative dynamical models that are
related to SOC models or have similar scaling laws, and discuss
what they have in common with SOC or where they differ. 
A metrics between processes and observables is synthesized
in Fig.~2.15. 

\subsection{Self-Organization Without Criticality (SO)} 

Self-organization (SO) is often referred to geometric patterns that 
originate by mutual interaction of their elements, without coordination 
from outside the system. For instance, convection
arranges itself into a regular pattern of almost equal-sized convection
cells, a structuring process also known as {\sl B\'enard cells}. Other
examples are the regular wavy pattern of sand dunes in the desert,
wavy patterns of Cirrus clouds in the Earth atmosphere,
Jupiter's atmosphere with white bands of ammonia ice clouds,
the spiral-like patterns of the Belousov-Zhabotinsky reaction-diffusion
system, or geometric patterns in biology, such as the skin of zebras,
giraffes, tigers, tropical fishes, or formation flight of birds.
From these examples we see that self-organization mostly refers to
the self-assembly of geometric patterns, which are more or less stable 
over long time intervals, although they arise from the physics of
non-equilibrium processes, which can involve diffusion, turbulence,
convection, or magneto-convection, which are governed by long-range 
interactions (via pressure and forces). 

What is the difference to self-organized criticality (SOC)? 
Are sandpile avalanches a self-organizing (SO) pattern? In the standard
scenario of the sandpile SOC model, individual avalanches are a local
phenomenon that are randomly triggered in space and time, but occur
independently, at least in the slowly-driven case. Thus, one avalanche
has no mutual interaction with another avalanche and the outcome of
the final size is independent of another, in contrast to self-organization
without criticality, where the interaction between system-wide structures
is coupled. In other words, sandpile avalanches are governed by localized
disturbances via next-neighbor interactions, while self-orga\-nizing
patterns may be formed by both next-neighbor and long-range interactions.
Another difference is that SO creates spatial patterns, while SOC
generates dynamical events (i.e., avalanches). Also the statistical
distributions of the two processes are different. A self-organizing
pattern is likely to produce a preferred size scale (e.g., the solar
granulation or the width of zebra stripes), while self-organized criticality
produces a scale-free powerlaw distribution of avalanche sizes.
The difference between SO and SOC could may be best illustrated with
a sandpile analogy. The same sandpile can be subject to self-organization
(SO), for instance when a steady wind blows over the surface and produces
wavy ripples with a regular spacing pattern, as well as be subject to
self-organized criticality (SOC), when intermittent avalanches occur
due to random-like disturbances by infalling sand grains. The former
geometric pattern may appear as a spatial-periodic pattern, while the
latter may exhibit a fractal geometry.

\subsection{Forced Self-Organized Criticality (FSOC)} 

In Bak's original SOC model, which is most genuinely reproduced by 
sandpile avalan\-ches and cellular automaton simulations, criticality
is continuously restored by next-neighbor interactions. If the local
slope becomes too step, an avalanche will erode it back to the critical
value. If the erosion by an avalanche flattened the slope too much,
it will be gradually restored by random input of dropped sand grains 
in the slowly-driven limit. When the observations were extended to 
magnetospheric substorms in the night-side geotail, powerlaw-like size 
distributions were found, suggesting a self-organized criticality process, 
but at the same time correlations with magnetic reconnection events at 
the dayside of the magnetosphere were identified, driven by
solar wind fluctuations, suggesting some large-scale transport 
processes and long-range coupling between the day-side and night-side
of the magnetosphere. Thus, these long-range interactions that
trigger local instabilities in the current sheet in the Earth's 
magnetotail were interpreted as external forcing or loading, and a combined
{\sl forced and/or self-organized criticality (FSOC)} process was suggested
(Chang 1992, 1998a,b, 1999a,b). A forced SOC process may apply to a
variety of SOC phenomena in the magnetosphere, such as 
magnetotail current disruptions (Lui et al.~2000),
substorm current disruptions (Consolini and Lui 1999),
bursty bulk flow events (Angelopoulos et al.~(1996, 1999),
magnetotail magnetic field fluctuations (Hoshino et al.~1994),
auroral UV blobs (Lui et al.~2000; Uritsky et al. 2002),
auroral optical blobs (Kozelov et al.~2004), 
auroral electron (AE) jets (Takalo et al.~1993; Consolini 1997),
or outer radiation belt electron events (Crosby et al.~2005).

What distinguishes the FSOC model from the standard (BTW-type) SOC model is 
mostly the long-range coupling, which is absent in sandpiles and cellular
automaton models. The driving force in the FSOC model is a long-distance 
action (loading process), while it is a random local disturbance in the 
BTW model. Otherwise, the FSOC model entails also powerlaw distributions for
the avalanche events, fractality, intermittency, statistical independence 
of events (indicated by random waiting time distributions), and a critical
threshold (for the onset of local plasma instabilities and/or magnetic
reconnection processes). 

\begin{figure}
\centerline{\includegraphics[width=0.9\textwidth]{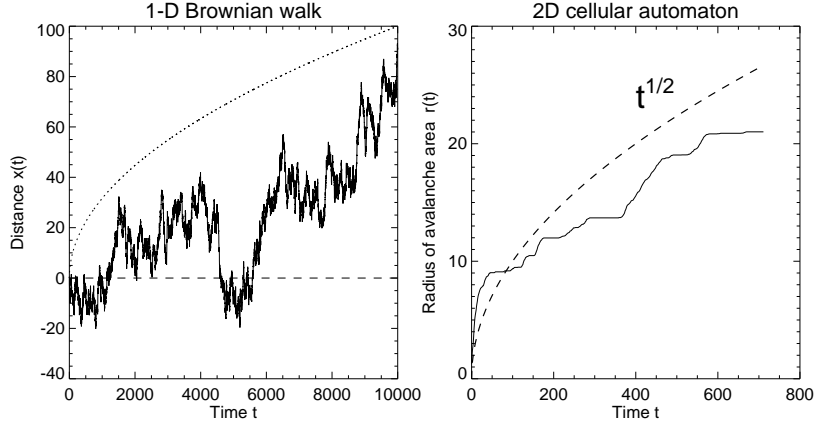}}
\caption{Numerical simulation of 1D Brownian random walk (left side)
and size increase of a 2D cellular automaton avalanche (right side)
as a function of time $t$. Both processes follow the trend
$x(t) \propto t^{1/2}$ expected for classical diffusion.}
\end{figure}

\subsection{Brownian Motion and Classical Diffusion} 

Brownian motion is a concept from classical physics that describes the
random motion of atoms or molecules in a gas (named after the Scottish
botanist Robert Brown), which can be best observed when a neutral gas of 
a different color is released in the atmosphere. What we observe then is an 
isotropic diffusion process (in the absence of forces or flows) that 
monotonically increases with the square-root of time, 
$$
	<x(t)> \propto t^{1/2} \ .
	\eqno(2.44)
$$
This statistical trend was derived in classical thermodynamics,
assuming a Gaussian distribution of velocities for the gas molecules,
so that the kinetic energy of particles follow a Boltzmann distribution
in thermodynamic equilibrium. The classical diffusion process can also
be described by a differential equation for a distribution function $f(x,t)$
of particles,
$$
	{\partial f(x,t) \over \partial t}
	=\kappa {\partial^2 f(x,t) \over \partial x^2} \ ,
	\eqno(2.45)
$$
which can describe heat transport, diffusion of gases, or magnetic diffusivity
on solar and stellar surfaces (i.e., manifested as meridional flows during a 
solar/stellar activity cycle).

An example of such a diffusive random
walk is simulated in Fig.~2.11 (left panel) for the 1-D variable $x(t)$. 
We have to be aware that a diffusive random walk $x(t)$ of a single 
particle can show large deviations from the expected trend $\propto t^{1/2}$,
which is only an expectation value for the statistical mean of many
random walks $<x(t)>=\sum_i x_i(t)$. In Fig.~11 (right panel) we show
also the time evolution of the mean radius $<r(t)>$ of a simulated 
cellular automaton avalanche (taken from Fig.~2.8, bottom right), which
shows the same trend of a time-dependence of $r(t) \propto t^{1/2}$.
Apparently, the enveloping volume of unstable cells in a SOC avalanche, 
defined by the
a mathematical re-distribution rule applied to a coarse-grained
lattice with a rough surface of stable, meta-stable, and unstable nodes
produces a similar time evolution as a random walk (as visualized
in Fig.~2.6), leading us to the fractal-diffusive SOC model described
in Section 2.2.2. This behavior of a diffusive transport process is
therefore common to both the Brownian motion (or classical diffusion)
and to a cellular automaton SOC process, but it operates for a SOC process
only during a finite time interval, as long as avalanche propagation is
enabled by finding unstable next neighbors, while a classical
diffusion process goes on forever without stopping. Therefore, we cannot
define an event and occurrence frequency distributions. However, we can
measure the fractal dimension of a random walk spatial pattern, which has
a Hausdorff dimension of $D_2=2$ in 2-D Euclidean space, or the power
spectrum, i.e., $P(\nu) \propto \nu^{-2}$, also called {\sl Brownian
noise}. 

\begin{figure}
\centerline{\includegraphics[width=1.\textwidth]{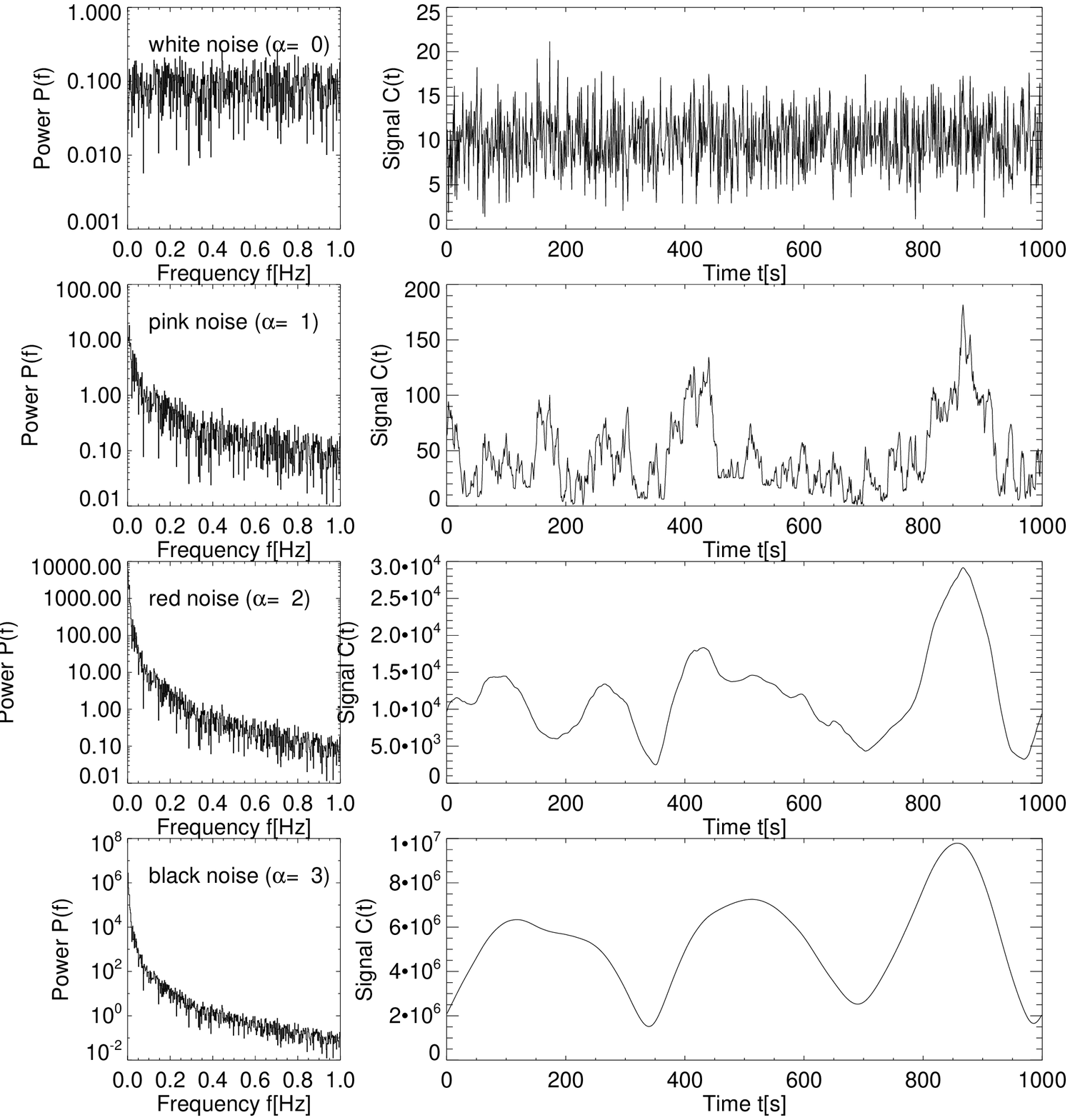}}
\captio{Noise power spectra (left panels) and corresponding time series
(right panels) for power spectral indices $p=0$ (top row: white noise
spectrum), $p=1$ (second row: pink noise spectrum),
$p=2$ (third row: red noise spectrum), and
$p=3$ (bottom row: black noise spectrum). The white noise
spectrum is multiplied with $\nu^{-p}$
in the other cases. The time series are reconstructed with the inverse
Fast Fourier Transform (Aschwanden 2011a).}
\end{figure}

\begin{table}
\begin{center}
\normalsize
\captio{Nomenclature of noise spectra.}
\begin{tabular}{lll}
\hline
\hline
Power spectrum            &Power index &Spectrum nomenclature  \\
\hline
$P(\nu) \propto \nu^{0}$  & $p=0$ & white noise \\

$P(\nu) \propto \nu^{-1}$ & $p=1$ & pink noise, flicker noise, $1/f$ noise \\
$P(\nu) \propto \nu^{-2}$ & $p=2$ & red noise, Brown(ian) noise \\
$P(\nu) \propto \nu^{-3}$ & $p=3$ & black noise         \\
\hline
\normalsize
\end{tabular}
\end{center}
\end{table}

Generalizations that include power spectra $P(\nu) \propto 
\nu^{-\beta}$ with arbitrary powerlaw indices $\beta$ have been dubbed
{\sl fractional Brownian motion (fBM)}, for instance white noise
($\beta=0$, where subsequent steps are uncorrelated), 1/f-noise, flicker
noise, or pink noise ($\beta=1$), Brownian noise or red noise ($\beta=2$),
or black noise ($\beta=3$, where subsequent time steps have some strong
correlations, producing slowly-varying time profiles). The latter 
fluctuation spectrum has been found to describe the stock market well.
Examples for these different types of fractional Brownian motion are
given in Fig.~2.12, while the nomenclature of noise spectra is summarized
in Table 2.3.

\subsection{Hyper-Diffusion and L\'evy Flight} 

The cellular automaton mechanism, the prototype of SOC dynamics,
involves a mathematical re-distribution rule amongst the next neighbor
cells (Eq.~2.1). This discretized numerical re-distribution rule has been
transformed in the continuum limit to an analytical function $A(x,t)$, which
can be expressed as fourth-order hyper-diffusion equation 
(Liu et al.~2002; Charbonneau et al.~2001),
$$
	{\partial A \over \partial t} 
	= -\kappa {\partial^4 A \over \partial x^4}
	= - {\partial^2 A \over \partial x^2}
	\kappa(A^2_{xx})
	{\partial^2 A \over \partial x^2} \ ,
$$
where $A$ represents the placeholder of the physical quantity corresponding 
to the symbol $z_{i,j,k}$ in Eq.~2.1, $x$ is the S-dimensional space coordinate
($x_{i,j,k}$ for $S=3$), with second-order centered differencing in space,
forward differencing in time, and $\kappa=1/2S$ the hyper-diffusion 
coefficient for Euclidean dimension $S$. However, in contrast to
classical physics, the hyper-diffusion coefficient $\kappa$ is subject
to a threshold value in the SOC model,
$$
	\kappa = \left\{
		\begin{array}{ll}
		\kappa_a \qquad	&{\rm if} \quad \Delta A^2 > A_{crit}^2 \\
		0               &{\rm otherwise} \\
		\end{array} 
		\right.  \ .
	\eqno(2.46)
$$
So, on a basic level, the
evolution of a cellular automaton avalanche translates into a 
hyper-diffusion process, as demonstrated in Liu et al.~(2002).
How can it be described by classical diffusion
as we discussed in the foregoing Section and in the fractal-diffusive
SOC model in Section 2.2.2? The major difference of the two apparently
contradicting descriptions lies in the fractality: The fourth-order
hyper-diffusion system (subjected to a threshold instability in the
diffusion coefficient $\kappa$) is applied to a Euclidean space with
dimension $S$, while the fractal-diffusive SOC model has a second-order 
(classical) diffusion coefficient in a fractal volume $V_S \propto x^{D_S}$ 
with a fractal dimension $D_S \approx (1+S)/2$. The two diffusion
descriptions apparently produce an equivalent statistics of avalanche
volumes $V_S$, as demonstrated by the numerical simulations of 
powerlaw distribution functions $N(V_S)$ by both models (Liu et al.~2002).

Other modifications of random walk or diffusion processes have been
defined in terms of the probability distribution function of step sizes.
Classical diffusion has a normal (Gaussian) distribution of step sizes,
which was coined {\sl Rayleigh flight} by Benoit Mandelbrot, while
{\sl L\'evy flight} was used for a heavy-tailed probability distribution
function (after the French mathematician Paul Pierre L\'evy).
Related are also the heavy-tailed Pareto probability distribution 
functions. Thus, L\'evy flight processes include occasional large-step
fluctuations on top of classical diffusion, which is found in
earthquake data, financial mathematics, cryptography, signal analysis,
astrophysics, biophysics, and solid state physics. A number of SOC 
phenomena have been
analyzed in terms of L\'evy flight processes, such as rice piles 
(Boguna and Corral 1997), random walks in fractal environments 
(Hopcraft et al.~1999; Isliker and Vlahos 2003), solar flare waiting 
time distributions (Lepreti et al.~2001), or extreme fluctuations 
in the solar wind (Moloney and Davidsen 2010, 2011).

\subsection{Nonextensive Tsallis Entropy} 

Related to L\'evy flight is also the {\sl nonextensive Tsallis entropy},
which originates from the standard extensive {\sl Boltzmann-Gibbs-Shannon
(BGS)} statistics, where the entropy $S$ is defined as,
$$
	S = -k_B \sum p_i \ln p_i \ ,
	\eqno(2.47)
$$
with $k_B$ the Boltzmann constant and $p_i$ the probabilities associated
with the microscopic configurations. The standard BGS statistics is
called {\sl extensive} when the correlations within the system are essentially
local (i.e., via next-neighbor interactions), while it is called
{\sl nonextensive} in the case when they are non-local or have long-range
coupling, similar to the local correlations in classical diffusion
and non-local steps in the L\'evy flights. As an example, the dynamic 
complexity of magnetospheric substorms and solar flares has been
described with nonextensive Tsallis entropy in Balasis et al.~(2011). 
Although nonlocal interactions are not part of the classical BTW automaton
model, the extensive Tsallis entropy produces statistical distributions
with similar fractality and intermittency. The nonextensive Tsallis
entropy has a variable parameter $q$,
$$
	S_q = k {1 \over q-1} \left( 1 - \sum_{i=1}^{W} p_i^q \right) \ ,
	\eqno(2.48)
$$
where $q$ is a measure of the nonextensivity of the system. A value of $q=1$
corresponds to the standard extensive BGS statistic, while a larger value
$q > 1$ quantifies the non-extensivity or importance of long-range coupling.
A value of $q=1.84$ was found to fit data of magnetospheric substorms 
and solar flare soft X-rays (Balasis et al.~2011).

\subsection{Turbulence} 

Turbulence is probably the most debated contender of SOC processes,
because of the many common observational signatures, such as the 
(scale-free) powerlaw distributions of spatial and temporal scales, 
the power spectra
of time profiles, random waiting time distributions, spatial fractality, 
and temporal intermittency. 

Let us first define turbulence. Turbulence was first defined in fluid 
dynamics in terms of the Navier-Stokes equation. A fluid is laminar at 
low Reynolds numbers (say at Reynolds numbers of $R \lapprox 5000$, 
defined by the dimensionless ratio of inertial to viscous forces, 
i.e., $R=v L /\nu_{visc}$ with $v$ the mean velocity of an object relative 
to the fluid, $L$ the characteristic linear dimension of the fluid, and 
$\nu_{vis}$ the kinematic viscosity), while it becomes turbulent at high 
Reynolds numbers. 

\begin{figure}
\centerline{\includegraphics[width=0.7\textwidth]{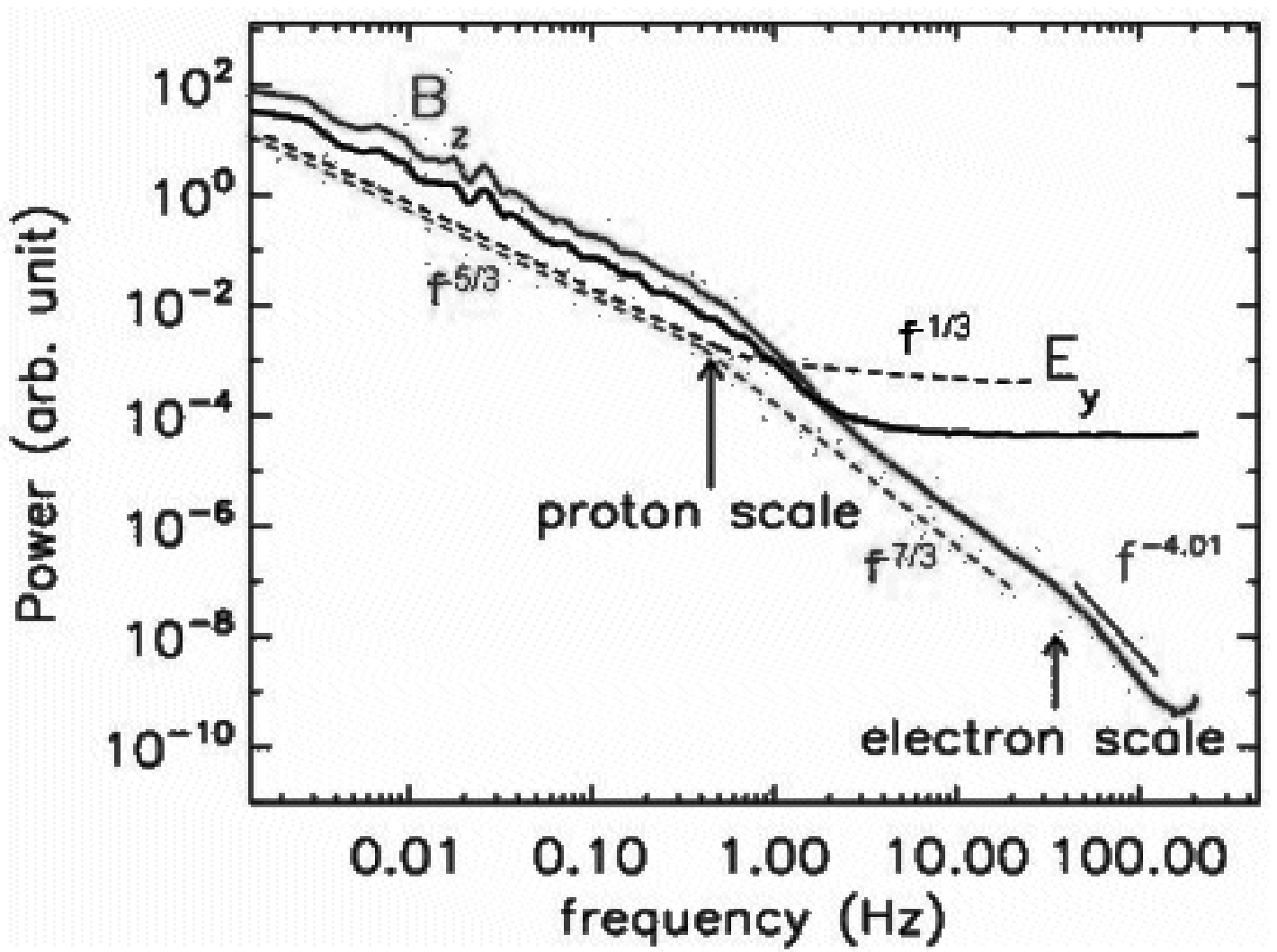}}
\captio{A spectrum of the solar wind is shown, based on {\sl CLUSTER} 
observations from large to small scales, with the proton and 
electron gyroradius scale indicated.
The solar wind spectrum is interpreted in terms of a turbulent MHD cascade,
with the theoretically predicted slopes of $f^{-5/3}$ and $f^{-7/3}$ from
gyro-kinetic theory. The plot proves that the energy continues cascading
below the proton scale down to the electron scale, where it is converted
to heat (via electron Landau damping resonance) causing the steepening of
the $B_z$ spectrum to $f^{-4}$ (Howes et al.~2008; Sahraoui et al.~2009;  
credit: ESA, CLUSTER).}
\end{figure}

SOC phenomena and turbulence have been studied in 
astrophysical plasmas (such as in the solar corona, in the solar wind, 
or in stellar coronae), where the magneto-hydrodynamic (MHD) behavior 
of the plasma is expressed with the continuity equation, the momentum
equation and induction equation,
$$
        \rho {D {\bf v} \over Dt}  = -\nabla p -\rho {\bf g} +
        ({\bf j} \times {\bf B})
        + {\nu}_{visc} \rho \left[ \nabla^2 {\bf v} + {1\over 3} \nabla
        (\nabla \cdot {\bf v}) \right]  \ ,
	\eqno(2.49)
$$
$$
        {\partial {\bf B} \over \partial t}
        = \nabla \times ({\bf v} \times {\bf B})
        + {\eta}_m \nabla^2 {\bf B} \ ,
	\eqno(2.50)
$$
with $\rho$ being the plasma density, $p$ the pressure, ${\bf B}$
the magnetic field, ${\bf j}$ the electric current density,
$\nu_{visc}$ the kinematic viscosity or shear viscosity,
$\eta_m=c^2/4\pi \sigma$ the magnetic diffusivity, and $\sigma$
the electric conductivity. In the solar corona and in the solar wind, 
the magnetic Reynolds number is sufficiently high ($R_m = v L /\eta_m \approx
10^8-10^{12}$) to develop turbulence. Turbulence in the coronal plasma or
in the solar wind is created by initial large-scale disturbances
(for instance by the convective random motion in sub-photospheric
layers or by the shock-like expansion of flares and coronal mass ejections).
The large-scale disturbances pump energy into the coronal magnetic field
or heliospheric solar wind at large scales, which cascade in the case
of turbulent flow to smaller scales, where the energy can be more
efficiently dissipated by friction, which is quantified by the kinematic
or (Braginskii) shear viscosity coefficient $\nu_{visc}$. Ultimately,
the energy of an MHD turbulent cascade is dissipated in the solar wind
at the spatial scale of (gyrating) thermal protons ($\approx 50$ km),
and at the scale of thermal electrons ($\approx 0.5$ km). The resulting
power spectrum of the solar wind shows a power spectrum of
$P(\nu) \propto \nu^{-5/3}$ at frequencies below the proton scale,
$P(\nu) \propto \nu^{-7/3}$ between the proton and electron scale,
and $P(\nu) \propto \nu^{-4}$ beyond the electron scale (Fig.~2.13).
For literature references on MHD turbulence in the solar corona and
in the solar wind see, e.g., Aschwanden (2011a, Section 10.4 therein).

What is the relationship between turbulence and SOC processes?
The transition from laminar flow to turbulent flow at a critical
Reynolds number $R_{crit}$ represents a similar thresholded instability 
criterion as the critical value $z_{crit}$ in SOC systems, above which
a turbulent avalanche could occur with subsequent cascading from 
vortices at large scales and little energy dissipation towards smaller 
scales with stronger energy dissipation. Turbulent energy dissipation
probably reduces the mean velocity of the fluid particles that are
faster than the background flow and laminar flow could be restored 
with lower particle velocities and a lower Reynolds number. To make this 
process self-organizing, we need also a driver mechanism that 
brings the fluid speed back up to the critical Reynolds number. 
In the solar wind, for instance, there is systematic acceleration
with heliocentric distance, which could drive the system from laminar
back to turbulent flows, and thus it could be self-organizing.
Fluctuations of the solar wind speed could therefore be 
considered as SOC avalanches. This could explain the scale-free
powerlaw distributions of spatial, temporal, and energy scales
measured in solar wind fluctuations, its fractality and intermittency.
If the critical threshold is exceeded only in localized regions, 
rather than in the entire system in a fully turbulent state, 
energy dissipation would also occur in locally unstable regions,
similar to SOC avalanches, and thus the two processes may exhibit
the same statistical distributions. Consequently, turbulence could
qualify as a SOC process if it is driven near the laminar-turbulent
critical Reynolds number in a self-organizing way. However, fully
developed turbulence, where the entire system is governed by a high 
(super-critical) Reynolds number, would correspond to a fast-driven
SOC system in a catastrophic phase with permanent avalanching, without
restoring the critical state as in a slowly-driven SOC system.

\subsection{Percolation} 

Percolation is a dynamic process that depends on the connectedness
and propagation probability of next-neighbor elements, and has a lot
in common with diffusion, fractal structures, and SOC avalanches.
A classical paradigm of a percolation process is coffee percolation,
which contains a solvent (water), a permeable substance (coffee grounds),
and soluble constituents (aromatic chemicals). The bimodal behavior of
a percolation process can be best expressed with the observation
whether a liquid that is poured on top of some porous material, will be
able to make its way from hole to hole and reach the bottom. The answer
will very much depend on the inhomogeneity characteristics of the
porous material. In the subcritical state, a percolating cluster will
exponentially die out, while it will propagate all the way to the bottom
in a supercritical state. The process can be described by branching
theory, where the connections between two next neighbors can be open
and let the liquid pass with probability $p$, or they can be closed
and the probability to pass is $(1-p)$. Completion of a pass from
the top to the bottom can be expressed as the combined statistical
probability along the entire pass. The critical value $p_{crit}$ that
decides between the two outcomes is $p_{crit}=1/2$ for a 2-D process.
So, this process has an extreme sensitivity to the system probability
$p$, depending whether it is below or above the critical value.
In SOC systems, the outcome of an avalanche is independent of the
initial conditions of the triggering disturbance. Also, the material
in a percolation process has a constant value of $p$ and does not
self-adjust to a critical value, and thus it does not behave like
a SOC system. However, what a percolation process has in common with
a SOC system is the fractality and intermittency of propagating features.
Other applications of percolation theory are used in physics, material 
science, complex networks, epidemiology, and geography.

\subsection{Phase Transitions} 

In classical thermodynamics, a phase transition describes the
transformation of one phase to another state of matter (solid,
liquid, gas, plasma). For instance, transitions between solid 
and liquid states are called ``freezing'' and ``melting'', between
liquid and gas are ``vaporization'' and ``condensation'',
between gas and plasma are ``ionization'' and ``deionization''.
Phase transitions from one state to another can be induced by
changing the temperature and/or pressure.

Since there are critical values of temperature and pressure
that demarcate the transitions, such as the freezing temperature
at $0^0$ C or the boiling point of $100^0$ C, we may compare
these critical limits with the critical threshold $z_{crit}$ in
SOC systems. Could this critical value be restored in a 
self-organizing way? To some extent, the daily weather changes
can be self-organizing. Let us assume some intermediate latitude
on our planet where the temperature is around the freezing point
at some seasonal period, say Colorado or Switzerland around November. 
The temperature may drop below the freezing point during night,
triggering snowfall, and may raise slightly above freezing point
during sunny days, triggering melting of snow. In this case, the
day-night cycle or the Earth's rotation causes the temperature to
oscillate around the freezing point, and therefore is in a 
self-organizing state. Additional variation of cloud cover may
introduce fluctuations on even shorter time scales. However,
this bimodal behavior of the temperature is not a robust operation
mode of self-organized criticality, because seasonal changes will
bring the average daily temperature systematically out of the
critical range around the freezing point. So, such temporary 
fluctuations around a critical point seem to be consistent with
a SOC system only on a temporary basis or in intermittent 
(seasonal) time intervals. 
In fact, a number of metereological phenomena have been found
to exhibit powerlaw distributions and where associated with SOC
behavior, such as Nile river fluctuations (Hurst 1951), rainfall
(Andrade et al.~1998), cloud formation (Nagel and Raschke 1992), 
climate fluctuations (Grieger 1992), aerosols in the atmosphere (Kopnin
et al.~2004), or forest fires (Kasischke and French 1995, Malamud 
et al.~1998). Phase transitions may be involved in some of these
processes (rainfall, cloud formation, forest fires).

Phase transitions are also involved in some laboratory experiments
in material or solid-state physics, quantum mechanics, and plasma physics,
that have been associated with SOC behavior, if we extend the term phase
transition also to morphological changes between highly-ordered and/or 
chaotic-structured patterns. Such phase transitions with SOC behavior
have been found in transitions between the ferromagnetic and paramagnetic 
phases, in specimens of ferromagnetic materials settling into one of 
a large number of metastable states that are not necessarily the 
energetically lowest ones (Che and Suhl 1990), in avalanche-like
topological rearrangements of cellular domain patterns in magnetic
garnet films (Babcock and Westervelt 1990), in the noise in the 
magnetic output of a ferromagnet when the magnetizing force applied 
to it is charged (Cote and Meisel 1991), called the Barkhausen effect,
in superconducting vortex avalanches in the Bean state (Field et al.~1995),
in tokamak plasma confinement near marginal stability driven by 
turbulence with a subcritical resistive pressure gradient (Carreras 
et al. 1996), or in the electrostatic floating potential fluctuations of
a DC glow charge plasma (Nurujjaman and Sekar-Iyenbgar 2007).

In summary, many SOC phenomena are observed during phase transitions,
producing avalanche-like rearrangements of geometric patterns, but
a phase-transition is not necessarily self-organizing in the sense
that the critical control parameter stays automatically tuned to the
critical point. A sandpile maintains its critical slope automatically
in a slowly-driven operation mode. If a thermostat would be available,
temperature-controlled phase transitions could be in a self-organized
operation mode, which indeed can be arranged in many situations.
So we conclude that phase transitions with some kind of ``thermostat''
tuned to the critical point represent indeed SOC systems. 

\begin{figure}
\centerline{\includegraphics[width=1.0\textwidth]{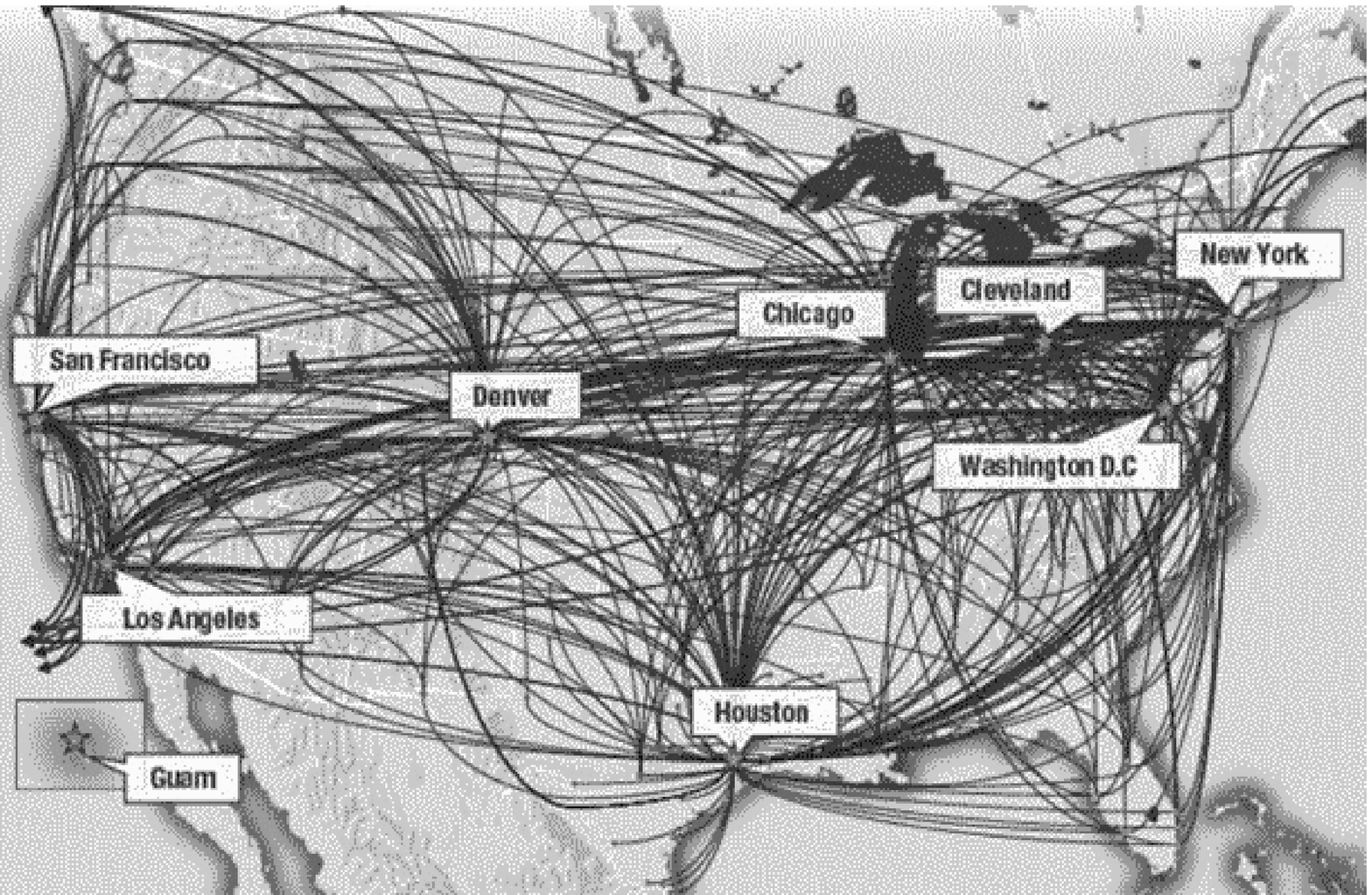}}
\captio{Network of American airline {\sl UNITED} with 8 US-based
hubs and connections to 378 destinations in 59 countries 
(Credit: United).}
\end{figure}

\subsection{Network Systems} 

The prototype of a SOC cellular automaton model is a regular lattice
grid, where re-distributions occur with next neighbors cells.
In contrast,
networks are irregular nets of nodes that are inter-connected in
manifold patterns, containing not necessarily only next-neighbor 
connections, but also arbitrary non-local, long-range connections.
Popular examples of networks are subway maps, city maps, road maps,
airplane route maps, electric grid maps, social networks, 
company organizational charts, financial networks, etc, all being
created by human beings in some way or another.
One big difference to SOC systems is therefore immediately clear,
the addition of long-range connections. Furthermore, while the
statistical probabilities for next-neighbor interactions are
identical for each lattice point in a SOC cellular automaton model,
networks can have extremely different connection probabilities 
at each node point. Nodes that have most of the connections are
also called hubs. For instance, the American airline {\sl United}
has 378 destinations, but server most of them from 10 hubs (Denver,
Washington DC, San Francisco, Los Angeles, Chicago, Houston, Cleveland,
Newark, Narita, Guam) (Fig.~14). Since the connection probability 
varies greatly from node to node, we do expect different clustering 
patterns in networks and in SOC lattice systems. 

Nevertheless, SOC behavior
has been studied in network systems like city growth and urban growth
(Zipf 1949; Zanette 2007), cotton prizes (Mandelbrot 1963),
financial stock market (Scheinkman and Woodford 1994; 
Sornette et al.~1996; Feigenbaum 2003; Bartolozzi et al.~2005), traffic jam
(Bak 1996; Nagel and Paczuski 1995), war casualties (Richardson 1941, 
Levy 1983), social networks (Newman et al.~2002), internet traffic
(Willinger et al. 2002), and language (Zipf 1949). SOC phenomena
with powerlaw-like frequency distributions have been observed for
phenomena that occur in these networks, measured by some quantity
that expresses fluctuations above some threshold or noise level.
Spatial patterns of network phenomena have been found to be fractal,
and time profiles of these fluctuations were found to be intermittent,
all hallmarks of SOC systems. So, are those network phenomena
self-organizing? What is the critical threshold and what drives the
system back to the critical threshold? The driving force 
of cotton prices, lottery wins, and stockmarket fluctuations is simply 
the human desire to gain money and to make profit. The driving mechanism for
traffic jams is certainly the daily need for commuting and transportation 
in a finite road network. The unpredictable number of casualties in
crimes and wars is a bit more controversial, but nobody would disagree
that the human desire for power and control plays a role. The driving
mechanism for social networks is most likely linked to curiosity and
desire for information. Since all these human traits are quite genuine
and persistent, a self-organizing behavior is warranted, where the
system is continuously driven towards a critical state, with occasional
larger fluctuations (Wall Street crash, Second World War, or other
``social catastrophes''), on top of the daily fluctuations on smaller
scales. In summary, most network-related phenomena can exhibit SOC
behavior, apparently with similar powerlaw-like statistics in grids
with non-uniform connectivities and non-local long-range connections
(in network systems) as in regular lattice grids with next-neighbor
interactions only (in classical SOC systems).  

\subsection{Chaotic Systems} 

Chaotic systems are nonlinear dissipative dynamical processes that 
are in principle deterministic, but exhibit ``chaotic'' behavior 
in the sense of irregular and fractal geometry, and intermittent 
time evolution. Some chaotic systems can be described as simple
as with two (such as the Lotka-Volterra equation) or three coupled
differential equations (e.g., the Lorenz equations). Classical
examples include forced pendulum, fluids near the onset of turbulence,
lasers, nonlinear optical devices, Josephson junctions, chemical 
reactions, the three-body system in celestial mechanics, ecological
population dynamics, or the heartbeat of biological organisms 
(e.g., Schuster 1988). If two
dynamical variables (say $x(t)$ and $y(t)$) are plotted in phase space
($y$ versus $x$), chaotic systems often exhibit a cyclic behavior,
with a limit cycle that is also called {\sl strange attractor}.
If a dynamical system is driven from small disturbances around
an equilibrium solution, where it exhibits a quasi-linear behavior,
bifurcations of system variables $x(t)$ can occur at critical values
with transitions to chaotic and intermittent behavior (also called
route to chaos). Such bifurcations ressemble the phase transitions
we discussed earlier (Section 2.3.8), which we generalized to
transitions from highly-ordered to chaotic-structured patterns.
The complexity of nonlinear chaotic systems is often characterized
with the {\sl dimension of strange attractors} (e.g., Grassberger
1985; Guckenheimer 1985), which approximately corresponds to the
number of coupled differential equations that is needed to describe
the system dynamics.

If we compare the dynamics of chaotic systems with SOC processes,
a major difference is the deterministic evolution of chaotic systems,
while SOC avalanches are statistically independent of each other,
even when they occur subsequently within a small (waiting) time interval.
A chaotic system evolves with a system-wide dynamical behavior that 
is influenced by the entire system, as the coupled differential 
equations express it, which includes next neighbors as well as
non-local, long-range coupling, while an avalanche in a SOC system
occurs only as a result of next-neighbor interactions. Do chaotic
systems have critical thresholds that are self-organizing? Chaotic
systems do have critical points, where onset of chaos starts, like at
pitchfork bifurcation point that leads to frequency doubling, but
there is usually not a driver that keeps the system at this critical
point. For instance, deterministic chaos sets in for fluids at the
transition from the laminar to the turbulent regime, but generally
there exists no driving mechanism that keeps a chaotic system
at the critical Reynolds number, so that the system is self-organizing
(Section 2.3.6). However, in other respects, chaotic systems exhibit
similar complexity as SOC systems, regarding fractality and intermittency,
and even powerlaw distributions may result in the statistics of chaotic
fluctuations. For weakly nonlinear dissipative systems near the limit
cycle, however, small fluctuations around a specific spatial scale or
temporal scale may produce Gaussian-like distributions, e.g., time
scales may be centered around the inverse frequency of the limit cycle.

Let us mention some examples of chaotic behavior in astrophysics.
Time series analysis of the X-ray variability of the neutron star
Her X-1 revealed a low-dimensional attractor (Voges et al.~1987; 
Cannizzo et al.~1990), as well as for the Vela pulsar (Harding et al.~1990).
Transient chaos was detected for the low-mass X-ray binary star
Scorpius X-1 (Scargle et al.~1993; Young and Scargle 1996), as well as
for the R scuti star (Buchler et al.~1996). The N-body system in
celestial mechanics can lead to chaotic behavior, such as for the
Saturnian moon Hyperion (Boyd et al.~1994). Chaotic behavior with
low-dimensional strange attractors and transitions to period doubling
was also found in solar radio bursts with quasi-periodic time series
(Kurths and Herzel 1986, Kurths and Karlicky 1989). Chaotic dynamics
was found in the solar wind (Polygiannakis and Moussas 1994),
in hydrodynamic convection simulations of the solar dynamo (Kurths
and Brandenburg 1991), and in solar cycle observations (Kremliovsky
1994; Charbonneau 2001; Spiegel 2009). Again, all these examples
reflect a system-wide nonlinear behavior, where subsequent fluctuations
are highly correlated in space and time, unlike the statistical
independency of SOC avalanches. However, chaotic systems may exhibit
similar statistics of fractal spatial and intermittent time scales
at critical points and transitions to chaotic dynamics, without
having an intrinsic mechanism that keeps the chaotic system near
this critical point in a self-organizing way. 

\begin{figure}
\centerline{\includegraphics[width=0.9\textwidth]{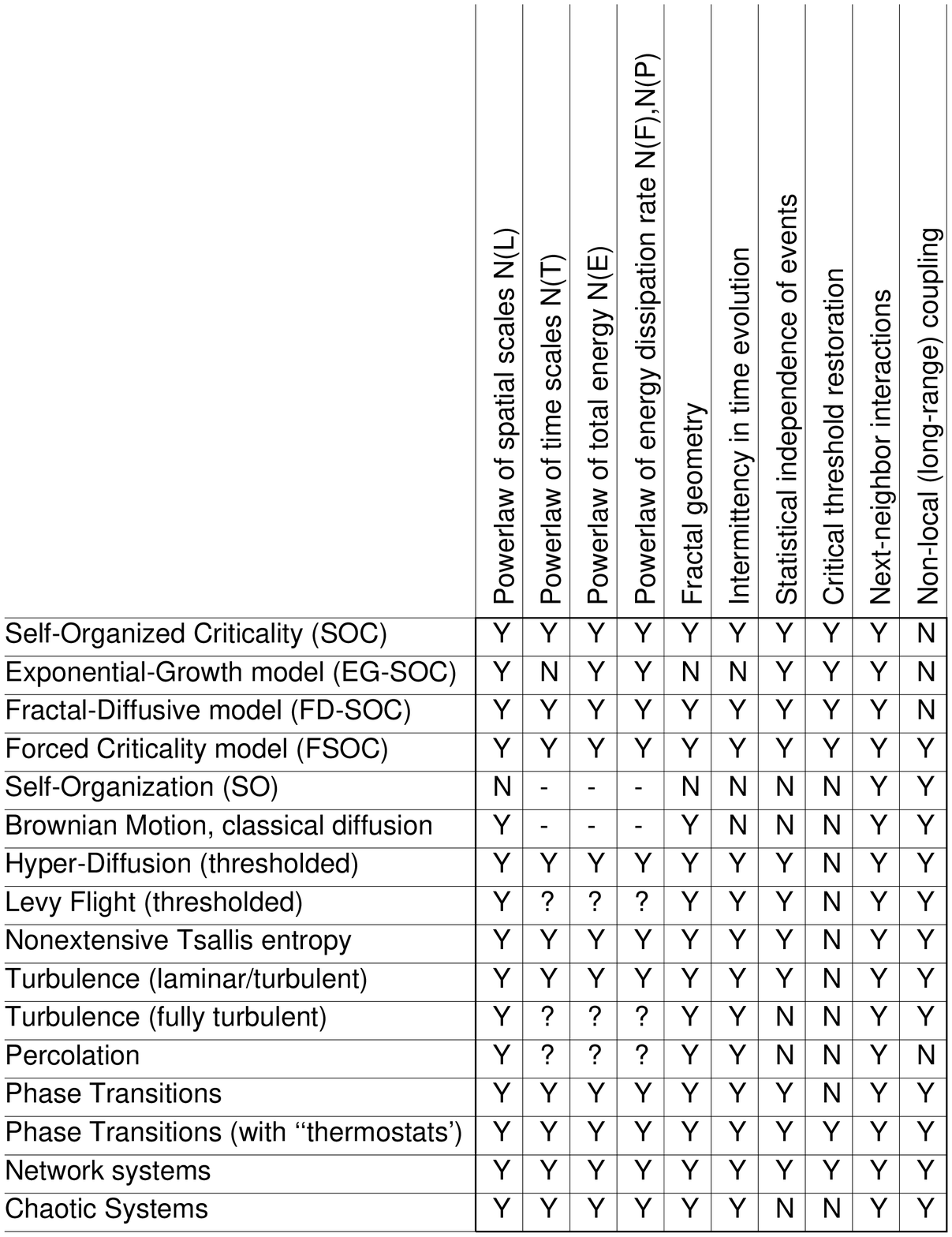}}
\captio{Metrics of SOC or SOC-like processes (rows) and observational
properties (columns), with an evaluation whether a specific process
can exhibit a particular observational characteristics (Y=Yes, N=No,
and the symbol $-$ is filled in for observational properties that do
not apply). See Chapter 2.3 for details.}
\end{figure}

\subsection{Synopsis} 

At the end of this chapter we summarize the characteristics of
SOC, SOC-related, and non-SOC processes in a metrics as shown
in Fig.~2.15. The SOC or SOC-like processes are listed in the
rows of Fig.~2.15, while the observational characteristics are
listed in the columns. The metrics tabulates whether a specific
SOC-like process can exhibit the main properties of SOC processes,
such as the powerlaw distributions of various parameters, the
fractal geometry, the temporal intermittency, the statistical
independence of events, the restoration of a critical threshold,
and next-neighbor interactions. The last column contains a 
property that SOC processes do not have, namely non-local and
long-range coupling, which may also represent an important
discrimination criterion between SOC and non-SOC processes,
a complementary characteristic to local and next-neighbor 
interaction processes such as the cellular automaton. The
properties listed in Fig.~2.15 reflect general trends rather than
strictly-valid matches. Many processes can exhibit powerlaw distributions
of parameters, but there exist always exceptions or deviations from
strict powerlaw distributions. The probably most fundamental
characteristics of SOC processes is a suitable mechanism that
restores the critical threshold for a instability, which often
is not automatically operating (such as in turbulence, phase transitions,
or chaotic systems), but can be artificially or naturally added to 
a process (such as a ``thermostat'' for phase transitions).  
Another important discrimination criterion between SOC and non-SOC
processes is the non-local or long-range coupling, which does not
exist in the classical SOC cellular automaton mechanism, but 
operates in many other processes (e.g., turbulence, phase transitions,
network systems, and chaotic systems), while the property of local
and next-neighbor interactions exists in virtually all processes, and 
thus is not discriminative. Finally, fractality, intermittency, and
powerlaw behavior is present in most of the processes, so it is
not a good discrimination criterion between SOC and non-SOC processes,
although powerlaws have always been considered as the hallmark of
SOC processes.

\section{References}

\ref{Andrade, R.F.S., Schellnhuber, H.J., and Claussen, M. 1998,
        {\sl Analysis of rainfall records: possible relation to self-organized
        criticality}, Physica A {\bf 254}, 3/4, 557-568.}
\ref{Angelopoulos, V., Coroniti, F.V., Kennel, C.F., Kivelson, M.G.,
        Walker, R.J., Russell, C.T., McPherron, R.L., Sanchez, E.,
        Meng, C.I., Baumjohann, W., Reeves, G.D., Belian, R.D., Sato, N.,
        Friis-Christensen, E., Sutcliffe, P.R., Yumoto,K., Harris,T. 1996,
        {\sl Multipoint analysis of a bursty bulk flow event on April 11,
        1985}, J. Geophys. Res. {\bf 101/A3}, 4967-4990.}
\ref{Angelopoulos, V., Mukai, T., and Kokubun, S. 1999,
        {\sl Evidence for intermittency in Earth's plasma sheet and
        implications for self-organized criticality},
        Phys. Plasmas {\bf 6/11}, 4161-4168.}
\ref{Aschwanden, M.J., Dennis, B.R., and Benz, A.O. 1998,
        {\sl Logistic avalanche processes, elementary time structures,
        and frequency distributions of flares},
        Astrophys. J. {\bf 497}, 972-993.}
\ref{Aschwanden, M.J. 2004,
        {\sl Physics of the Solar Corona. An Introduction},
	ISBN 3-540-22321-5, PRAXIS, Chichester, UK, and Springer, Berlin, 842p.}
\ref{Aschwanden, M.J. 2011a,
 	{\sl Self-Organized Criticality in Astrophysics. The Statistics 
	of Nonlinear Processes in the Universe}, ISBN 978-3-642-15000-5, 
	Springer-Praxis: New York, 416p.}
\ref{Aschwanden, M.J. 2011b,
 	{\sl The state of self-organized criticality of the Sun during 
	the last 3 solar cycles. I. Observations},
 	Solar Phys. {\bf 274}, 99-117.}
\ref{Aschwanden, M.J. 2012,
 	{\sl A statistical fractal-diffusive avalanche model of a 
	slowly-driven self-organized criticality system},
 	Astron.Astrophys. {\bf 539}, A2 (15 p).}
\ref{Aschwanden, M.J. and Freeland, S.L. 2012,
 	{\sl Automated solar flare statistics in soft X-rays over 37 years 
	of GOES observations - The Invariance of self-organized criticality 
	during three solar cycles}, Astrophys. J. (subm).}
\ref{Babcock, K.L. and Westervelt, R.M. 1990,
        {\sl Avalanches and self-organization in cellular magnetic-domain
        patterns}, Phys. Rev. Lett. {\bf 64/18}, 2168-2171.}
\ref{Bak, P., Tang, C., and Wiesenfeld, K. 1987,
        {\sl Self-organized criticality - An explanation of 1/f noise},
        Physical Review Lett. {\bf 59/27}, 381-384.}
\ref{Bak, P., Tang, C., and Wiesenfeld, K. 1988,
        {\sl Self-organized criticality},
        Physical Rev. A {\bf 38/1}, 364-374.}
\ref{Bak, P. and Chen, K. 1989,
        {\sl The physics of fractals},
        Physica D {\bf 38}, 5-12.}
\ref{Bak, P., Chen, K. and Creutz, M. 1989,
        {\sl Self-organized criticality in the ``Game of Life''},
        Nature {\bf 342}, 780-781.}
\ref{Bak, P. and Sneppen, K. 1993,
        {\sl Punctuated equilibrium and criticality in a simple model
        of evolution}, Phys. Rev. Lett. {\bf 71/24}, 4083-4086.}
\ref{Bak, P. 1996,
        {\sl How nature works},
        Copernicus, Springer-Verlag, New York.}
\ref{Balasis, G., Daglis, I.A., Anastasiadis, A., Papadimitriou, C., 
	Mandea, M., and Eftaxias,K. 2011,
 	{\sl Universality in solar flare, magnetic storm and earthquake 
	dynamics using Tsallis statistical mechanics},
 	Physica A {\bf 390}, 341-346.}
\ref{Bartolozzi, M., Leinweber, D.B., and Thomas, A.W. 2005,
        {\sl Self-organized criticality and stock market dynamics:
        an empirical study}, Physica A {\bf 350}, 451-465.}
\ref{Boguna, M. and Corral, A. 1997,
 	{\sl Long-tailed trapping times and L\'evy flights in a 
	self-organized critical granular system},
	Phys.Rev.Lett. {\bf 78/26}, 4950-4953.}
\ref{Boyd, P.T., Mindlin, G.B., Gilmore, R., and Solari, H.G. 1994,
        {\sl Topological analysis of chaotic orbits: revisiting Hyperion},
        Astrophys. J. {\bf 431}, 425-431.}
\ref{Buchler, J.R., Kollath, Z., Serre, T., and Mattei, J. 1996,
        {\sl Nonlinear analysis of the light curve of the variable star
        R Scuti}, Astrophys. J. {\bf 462}, 489-501.}
\ref{Burridge R. and Knopoff L. 1967,
        {\sl Model and theoretical seismicity},
        Seis. Soc. Am. Bull. {\bf 57}, 341-347.}
\ref{Cannizzo, J.K., Goodings, D.A., and Mattei, J.A. 1990,
        {\sl A search for chaotic behavior in the light curves of three
        long-term variables}, Astrophys. J. {\bf 357}, 235-242.}
\ref{Carreras, B.A., Newman, D., Lynch, V.E., and Diamond, P.H., 1996,
        {\sl A model realization of self-organized criticality for plasma
        confinement}, Phys. Plasmas {\bf 3/8}, 2903-2911.}
\ref{Chang, T.S. 1992,
        {\sl Low-dimensional behavior and symmetry breaking of stochastic
        systems near criticality - Can these effects be observed in space
        and in the laboratory},
        IEEE Trans. Plasma Sci. {\bf 20/6}, 691-694.}
\ref{Chang, T.S. 1998a,
        {\sl Sporadic, Localized reconnections and mnultiscale intermittent
        turbulence in the magnetotail}, in {\sl Geospace Mass and Energy
        Flow} (eds. Horwitz, J.L., Gallagher, D.L., and Peterson, W.K.),
        AGU Geophysical Monograph {\bf 104}, p.193.}
\ref{Chang, T.S. 1998b,
        {\sl Multiscale intermittent turbulence in the magnetotail},
        in {\sl Proc. 4th Intern. Conf. on Substorms}, (eds. Kamide, Y. et al.),
        Kluwer Academic Publishers, Dordrecht, and Terra Scientific
        Company, Tokyo, p.431.}
\ref{Chang, T.S. 1999a,
        {\sl Self-organized criticality, multi-fractal spectra, and
        intermittent merging of coherent structures in the magnetotail},
        Astrophys. Space Sci. {\bf 264}, 303-316.}
\ref{Chang, T.S. 1999b,
        {\sl Self-organized criticality, multi-fractal spectra, sporadic
        localized reconnections and intermittent turbulence in the
        magnetotail}, Phys. Plasmas {\bf 6/11}, 4137-4145.}
\ref{Charbonneau, P., McIntosh, S.W., Liu, H.L., and Bogdan, T.J. 2001,
        {\sl Avalanche models for solar flares},
        Solar Phys. {\bf 203}, 321-353.}
\ref{Che, X. and Suhl, H., 1990,
        {\sl Magnetic domain pattern as self-organizing critical systems},
        Phys. Rev. Lett. {\bf 64/14}, 1670-1673.}
\ref{Consolini, G. 1997,
        {\sl Sandpile cellular automata and magnetospheric dynamics},
        in Proc, Cosmic Physics in the year 2000, (eds. S.Aiello,
        N.Iucci, G.Sironi, A.Treves, and U.Villante), SIF: Bologna, Italy,
        Vol. {\bf 58}, 123-126.}
\ref{Consolini, G. and Lui, A.T.Y. 1999,
        {\sl Sign-singularity analysis of current disruption},
        Geophys. Res. Lett. {\bf 26/12}, 1673-1676.}
\ref{Cote, P.J. and Meisel, L.V. 1991,
        {\sl Self-organized criticality and the Barkhausen effect},
        Phys. Rev. Lett. {\bf 67}, 1334-1337.}
\ref{Crosby, N.B., Meredith, N.P., Coates, A.J., and Iles, R.H.A. 2005,
        {\sl Modelling the outer radiation belt as a complex system in a
        self-organised critical state},
        Nonlinear Processes in Geophysics {\bf 12}, 993-1001.}
\ref{Crosby, N.B. 2011,
 	{\sl Frequency distributions: From the Sun to the Earth},
 	Nonlinear Processes in Geophysics {\bf 18/6}, 791-805.}
\ref{Dennis, B.R. and Zarro, D.M. 1993,
        {\sl The Neupert effect: what can it tell us about the impulsive
        and gradual phases of solar flares},
        Solar Phys. {\bf 146}, 177-190.}
\ref{Dennis, B.R., Veronig, A., Schwartz, R.A., Sui, L., Tolbert, A.K., 
	Zarro,D.M. and the RHESSI Team 2003,
 	{\sl The Neupert effect and new RHESSI measures of the total 
	energy in electrons accelerated in solar flares},
 	Adv.Space Res. {\bf 32}, 2459.}
\ref{Feigenbaum, J.A. 2003,
        {\sl Financial physics},
        Rep. Prog. Phys. {\bf 66}, 1611-1649.}
\ref{Fermi, E. 1949,
        {\sl On the origin of the cosmic radiation},
        Phys. Rev. Lett. {\bf 75}, 1169-1174.}
\ref{Field, S., Witt, J., Nori, F., and Ling, X. 1995,
        {\sl Superconducting vortex avalanches},
        Phys. Rev. Lett.  {\bf 74}, 1206-1209.}
\ref{Grassberger, P. 1985,
        {\sl Information aspects of strange attractors},
        in {\sl Chaos in Astrophysics} (eds. Buchler, J.R. et al.),
        Reidel Publishing Company: Dordrecht, p.193-222.}
\ref{Grieger, B. 1992,
        {\sl Quaternary climatic fluctuations as a consequence of
        self-organized criticality},
        Physica A {\bf 191},  51-56.}
\ref{Guckenheimer J. 1985,
        {\sl Clues to strange attractors},
        in {\sl Chaos in Astrophysics} (eds. Buchler, J.R. et al.),
        Reidel Publishing Company: Dordrecht, p.185-191.}
\ref{Harding, A.K., Shinbrot, T., and Cordes, J.M. 1990,
        {\sl A chaotic attractor in timing noise from the Vela pulsar?},
        Astrophys. J. {\bf 353}, 588-596.}
\ref{Hergarten, S. and Neugebauer, H.J. 1998,
        {\sl Self-organized criticality in a sandslide model},
        Geophys. Res. Lett. {\bf 25/6}, 801-804.}
\ref{Hopcraft, K.I., Jakeman, E., and Tanner, R.M.J. 1999,
 	{\sl Levy random walks with fluctuating step number and multiscale 
	behavior}, Phys.Rev E {\bf 60/5}, 5327-5343.}
\ref{Hoshino, M., Nishida, A., Yamamoto, T., and Kokubun, S. 1994,
        {\sl Turbulent magnetic field in the distant magnetotail:
        bottom-up process of plasmoid formation?},
        Geophys. Res. Lett. {\bf 21/25}, 2935-2938.}
\ref{Howes, G.G., Dorland, W., Cowley, S.C., Hammett, G.W., Quataert, E.,
        Schekochihin, A.A., and Tatsuno, T. 2008,
        {\sl Kinetic simulations of magnetized turbulence in astrophysical
        plasmas}, Phys. Rev. Lett. {\bf 100/6}, 065004.}
\ref{Huberman, B.A. and Adamic, L. 1999,
        {\sl Growth dynamics of the World-Wide Web},
        Nature {\bf 401}, 131.}
\ref{Hurst, H.E. 1951,
        {\sl Long-term storage capacity of reservoirs},
        Trans. Am. Soc. Civil Eng. {\bf 116}, 770-799.}
\ref{Isliker, H., Anastasiadis, A., Vassiliadis, D., and Vlahos, L. 1998,
        {\sl Solar flare cellular automata interpreted as discretized
        MHD equations}, Astron. Astrophys. {\bf 335}, 1085-1092.}
\ref{Isliker, H. and Vlahos, L. 2003,
	{\sl Random walk through fractal environments},
	Phys. Rev. E, 67/2, id. 026413.}
\ref{Jensen, H.J. 1998,
        {\sl Self-Organized Criticality. Emergent complex behavior in
        physical and biological systems},
        Cambridge University Press, Cambridge UK, 153 p.}
\ref{Kasischke, E.S.and French N.H.F. 1995,
        Remote Sens. Environ. {\bf 51}, 263-275.}
\ref{Kopnin, S.I., Kosarev, I.N., Popel, S.I., and hyu, M.Y. 2004,
        {\sl Localized structures of nanosize charged dust in Earth's
        middle atmosphere}, Planet. Space Sci. {\bf 52/13}, 1187-1194.}
\ref{Kozelov, B.V., Uritsky, V.M., and Klimas, A.J. 2004,
        {\sl Power law probability distributions of multiscale auroral
        dynamics from ground-based TV observations},
        Geophys. Res. Lett. {\bf 31/20}, CiteID L20804.}
\ref{Kremliovsky, M.N. 1994,
        {\sl Can we understand time scales of solar activity?},
        Solar Phys. {\bf 151}, 351-370.}
\ref{Kurths, J. and Herzel, H. 1986,
        {\sl Can a solar pulsation event be characterized by a
        low-dimensional chaotic attractor},
        Solar Phys. {\bf 107}, 39-45.}
\ref{Kurths, J. and Karlicky, M. 1989,
        {\sl The route to chaos during a pulsation event},
        Solar Phys. {\bf 119}, 399-411.}
\ref{Kurths, J. and Brandenburg, A. 1991,
        {\sl Lyapunov exponents for hydrodynamic convection},
        Phys. Rev. A. {\bf 44/6}, 3427-3429.}
\ref{Lepreti, F., Carbone, V., and Veltri, P. 2001,
        {\sl Solar flare waiting time distribution: varying-rate Poisson
        or Levy function?}, Astrophys. J. {\bf 555}, L133-L136.}
\ref{Levy, J.S. 1983,
        {\sl War in the Modern Great Power System 1495-1975},
        (Lexington, KY: University of Kentucky Press), p. 215.}
\ref{Liu, H., Charbonneau, P., Pouquet, A., Bogdan, T., and McIntosh, S.W.
        2002, {\sl Continuum analysis of an avalanche model for solar flares},
        Phys. Rev. E {\bf 66}, 056111.}
\ref{Lu, E.T. and Hamilton, R.J. 1991,
        {\sl Avalanches and the distribution of solar flares},
        Astrophys. J. {\bf 380}, L89-L92.}
\ref{Lui, A.T.Y., Chapman, S.C., Liou,K., Newell, P.T., Meng, C.I.,
        Brittnacher, M., and Parks, G.K. 2000,
        {\sl Is the dynamic magnetosphere an avalanching system?},
        Geophys. Res. Lett. {\bf 27/7}, 911-914.}
\ref{Malamud, B.D., Morein, G., and Turcotte D.L. 1998,
        {\sl Forest fires: An example of self-organized critical behavior},
        Science {\bf 281}, 1840-1842.}
\ref{Mandelbrot, B.B. 1963,
        {\sl The variation of certain speculative prices},
        Journal of Business of the University of Chicago {\bf 36/4}, 394-419.}
\ref{Mineshige, S., Ouchi,N.B., and Nishimori, H. 1994a,
        {\sl On the generation of 1/f fluctuations in X-rays from
        black-hole objects}, Publ. Astron. Soc. Japan {\bf 46}, 97-105.}
\ref{Mineshige, S., Takeuchi, M., and Nishimori, H. 1994b,
        {\sl Is a black hole accretion disk in a self-organized critical
        state ?}, Astrophys. J. {\bf 435}, L125-L128.}
\ref{Moloney, N.R. and Davidsen, J. 2010,
 	{\sl Extreme value statistics in the solar wind: An application 
	to correlated Levy processes}, JGR (Space Physics), {\bf 115}, 10114.}
\ref{Moloney, N.R. and Davidsen, J. 2011,
 	{\sl Extreme bursts in the solar wind},
 	GRL {\bf 381}, 14111.}
\ref{Morales, L. and Charbonneau, P. 2008,
        {\sl Self-organized critical model of energy release in an idealized
        coronal loop}, Astrophys. J. {\bf 682}, 654-666.}
\ref{Nagel, K. and Raschke, E. 1992,
        {\sl Self-organized criticality in cloud formation?},
        Physica A {\bf 182/4}, 519-531.}
\ref{Nagel, K. and Paczuski, M. 1995,
        {\sl Emergent traffic jams}, Phys. Rev. E. {\bf 51/4}, 2909-2918.}
\ref{Newman, M.E.J., Watts, D.J., and Strogatz, S.H. 2002,
        {\sl Random graph models of social networks},
        in {\sl Self-organized complexity in the physical, biological,
        and social sciences}, Arthur M. Sackler Colloquia,
        (eds. Turcotte, D., Rundle, J., and Frauenfelder, H.),
        The National Academy of Sciences: Washington DC, p.2566-2572.}
\ref{Nurujjaman, Md. and Sekar-Iyenbgar, A.N. 2007,
        {\sl Realization of SOC behavior in a DC glow discharge plasma},
        Phys. Lett. A {\bf 360},  717-721.}
\ref{Pavlidou, V., Kuijpers, J., Vlahos, L., and Isliker, H. 2001,
        {\sl A cellular automaton model for the magnetic activity in
        accretion disks}, Astron. Astrophys. {\bf 372}, 326-337.}
\ref{Polygiannakis, J.M. and Moussas, X. 1994,
        {\sl On experimental evidence of chaotic dynamics over short time
        scales in solar wind and cometary data using nonlinear prediction
        techniques}, Solar Phys. {\bf 151}, 341-350.}
\ref{Pruessner, G. 2012,
 	{\sl Self-organised criticality. Theory, models and characterisation},
 	ISBN 978-0-521-85335-4, Cambridge University Press, Cambridge.}
\ref{Reed, W.J. and Hughes, B.D. 2002,
        {\sl From gene families and genera to incomes and internet file sizes:
        Why power laws are so common in nature},
        Phys. Rev. Lett. E {\bf 66}, 067103.}
\ref{Richardson, L.F. 1941,
        {\sl Frequency occurrence of wars and other fatal quarrels},
        Nature {\bf 148}, 598.}
\ref{Rosner, R., and Vaiana, G.S. 1978,
        {\sl Cosmic flare transients: constraints upon models for energy 
	storage and release derived from the event frequency distribution},
        Astrophys. J. {\bf 222}, 1104-1108.}
\ref{Sahraoui, F., Goldstein, M.L., Robert, P., Khotyzintsev, Y.V. 2009,
        {\sl Evidence of a cascade and dissipation of solar-wind turbulence
        at the electron gyroscale}, Phys. Rev. Lett. {\bf 102}, 231102:1-4.}
\ref{Scargle, J. 1993,
        {\sl Wavelet methods in astronomical time series analysis},
        Internat. Conf. on {\sl Applications of time series analysis in
        astronomy and meteorology}, (ed. O. Lessi), Padova, Italy.}
\ref{Scheinkman, J.A. and Woodford, M. 1994,
        {\sl Self-organized criticality and economic fluctuations},
        Am. Econ. Rev. {\bf 84}, 417-421.}
\ref{Schuster, H.G. 1988,
        {\sl Deterministic Chaos: An Introduction},
        Weinheim (Germany): VCH Verlagsgesellschaft, 270 p.}
\ref{Sornette, D. 2004,
        {\sl Critical phenomena in natural sciences: chaos, fractals,
        self-organization and disorder: concepts and tools},
        Springer, Heidelberg, 528 p.}
\ref{Spiegel, E.A. 2009,
        {\sl Chaos and Intermittency in the Solar Cycle},
        Space Sci. Rev. {\bf 144}, 25-51.}
\ref{Sornette, D., Johansen, A., and Bouchard, J.P. 1996,
        {\sl Stock market crashes, precursors and replicas},
        J. Physique I {\bf 6}, 167-175.}
\ref{Takalo, J. 1993,
        {\sl Correlation dimension of AE data},
        Ph. Lic. Thesis, {\sl Laboratory report 3}, Dept. Physics,
        University of Jyv\"askyl\"a.}
\ref{Takalo, J., Timonem, J., Klimas, A., Valdivia, J., and Vassiliadis, D.
        1999a, {\sl Nonlinear energy dissipation in a cellular automaton
        magnetotail field model}
        Geophys. Res. Lett. {\bf 26/13}, 1813-1816.}
\ref{Takalo, J., Timonem, J., Klimas, A., Valdivia, J., and Vassiliadis, D.
        1999b, {\sl A coupled-map model for the magnetotail current sheet},
        Geophys. Res. Lett. {\bf 26/19}, 2913-2916.}
\ref{Turcotte, D.L. 1999,
        {\sl Self-organized criticality},
        Rep. Prog. Phys. {\bf 62}, 1377-1429.}
\ref{Uritsky, V.M., Klimas, A.J., Vassiliadis, D., Chua, D., and Parks, G.
        2002, {\sl Scale-free statistics of spatiotemporal auroral emission
        as depicted by Polar UVI images: dynamic magnetosphere is an
        avalanching system}, J. Geophys. Res. {\bf 1078/A12},
        SMP 7-1, CiteID 1426.}
\ref{Vassiliadis, D., Anastasiadis, A., Georgoulis, M., and Vlahos L. 1998,
        {\sl Derivation of solar flare cellular automata models from a
        subset of the magnetohydrodynamic equations},
        Astrophys. J. {\bf 509}, L53-L56.}
\ref{Vlahos, L., Georgoulis, M., Kliuiving, R., and Paschos, P. 1995,
        {\sl The statistical flare}, Astron. Astrophys. {\bf 299}, 897-911.}
\ref{Voges, W., Atmanspacher, H., and Scheingraber, H. 1987,
        {\sl Deterministic chaos in accretion systems: Analysis of the X-ray
        variability of Hercules X-1},
        Astrophys. J. {\bf 320}, 794-802.}
\ref{Willinger W., Govindan, R., Jamin, S., Paxson, V., and Shenker, S. 2002,
        {\sl Scaling phenomena in the Internet: critically examining
        criticality},
        in {\sl Self-organized complexity in the physical, biological,
        and social sciences}, Arthur M. Sackler Colloquia,
        (eds. Turcotte, D., Rundle, J., and Frauenfelder, H.),
        The National Academy of Sciences: Washington DC, p.2573-2580.}
\ref{Willis, J.C. and Yule, G.U. 1922,
        {\sl Some statistics of evolution and geographical distribution
        in plants and animals, and their significance},
        Nature {\bf 109}, 177-179.}
\ref{Wolfram, S. 2002,
        {\sl A new kind of science},
        Wolfram Media, ISBN 1-57955-008-8.}
\ref{Yoder, M.R., Holliday, J.R., Turcotte, D.L., and Rundle,J.B. 2012,
 	{\sl A geometric frequency-magnitude scaling transition: Measuring 
	b=1.5 for large earthquakes}, Tectonophysics {\bf 532-535}, 167-174.}
\ref{Young, K. and Scargle, J.D. 1996,
        {\sl The dripping handrail model: Transient chaos in accretion
        disks}, Astrophys. J. {\bf 468}, 617-632.}
\ref{Zanette, D.H. 2007,
        {\sl Multiplicative processes and city sizes},
        in  "The Dynamics of Complex Urban Systems. An Interdisciplinary
        Approach",  (eds. S.  Albeverio, D. Andrey, P.  Giordano, and
        A. Vancheri, eds. (Springer, Berlin, 2007).}
\ref{Zipf, G.K. 1949,
        {\sl Human Behavior and the Principle of Least Effort},
        Cambridge MA, Addison-Wesley.}